\documentclass[11pt]{article}

\usepackage{graphicx}
\usepackage{bm}
\usepackage{siunitx}
\usepackage{booktabs}
\usepackage{xcolor}
\usepackage{authblk}
\usepackage{amsmath}
\usepackage{amssymb}
\usepackage{amsthm}
\usepackage{multirow}
\usepackage{caption}
\usepackage{todonotes}

\usepackage{epstopdf}

\usepackage{pdfpages}
\usepackage{algorithm} 
\usepackage{algpseudocode}

\usepackage{stackengine}
\newcommand\xrowht[2][0]{\addstackgap[.5\dimexpr#2\relax]{\vphantom{#1}}}

\usepackage{float}
\usepackage{placeins} 
\usepackage{commath}
\usepackage{fancybox}
\usepackage{tikz}
\usepackage[hidelinks]{hyperref}
\usepackage{url}
\usepackage[margin=0.5in]{geometry}
\usepackage[toc,page]{appendix}

\usepackage[backend=biber, style=authoryear, citestyle=authoryear, autocite=inline, isbn=true, maxnames=6]{biblatex} 

\DeclareCiteCommand{\hciteauthor}
  {\usebibmacro{prenote}}
  {\printtext[bibhyperref]{\printnames{labelname}\iffieldundef{year}{}{ (\printfield{year})}}}
  {\multicitedelim}
  {\usebibmacro{postnote}}

\DeclareCiteCommand{\hcite}
  {\usebibmacro{prenote}}
  {\printtext[bibhyperref]{\printnames{labelname}\iffieldundef{year}{}{\addspace\printfield{year}}}}
  {\multicitedelim}
  {\usebibmacro{postnote}}

\definecolor{lime}{HTML}{A6CE39}

\DeclareRobustCommand{\orcidicon}{
	\begin{tikzpicture}
	\draw[lime, fill=lime] (0,0) 
	circle [radius=0.16] 
	node[white] {{\fontfamily{qag}\selectfont \tiny ID}};
	\draw[white, fill=white] (-0.0625,0.095) 
	circle [radius=0.007];
	\end{tikzpicture}
	\hspace{-2mm}
}
\foreach \x in {A, ..., Z}{\expandafter\xdef\csname orcid\x\endcsname{\noexpand\href{https://orcid.org/\csname orcidauthor\x\endcsname}
			{\noexpand\orcidicon}}
}

\newcommand{\orcid}[1]{\href{https://orcid.org/#1}{\textcolor[HTML]{A6CE39}{\orcidicon}}}

\theoremstyle{definition}
\newtheorem{Def}{Definition}[section]

\newtheorem{Thm}[Def]{Theorem}

\NewBibliographyString{diplomathesis}
\DefineBibliographyStrings{english}{
  diplomathesis = {Graduation Project Report. National Engineering Diploma},
}

\NewBibliographyString{masterthesis}
\DefineBibliographyStrings{english}{
  masterthesis = {Master of Science Degree Thesis. King Abdullah University of Science and Technology},
}

\begin{document}

\title{Data-driven uncertainty quantification for constrained stochastic differential equations and application to solar photovoltaic power forecast data}  

\author[1]{Khaoula Ben Chaabane\orcid{0000-0001-8299-0703}}
\author[2]{Ahmed Kebaier\orcid{0000-0003-3468-8811}}
\author[3]{Marco Scavino\orcid{0000-0001-5114-853X}}
\author[4, 5]{ Ra\'ul  Tempone\orcid{0000-0003-1967-4446}}

\affil[1]{Chair of Mathematics for Uncertainty Quantification, RWTH Aachen University, Germany}
\affil[2]{Laboratoire de Mathématiques et Modélisation d'\'Evry, CNRS, UMR 8071, Univ \'Evry, Université Paris-Saclay, \'Evry, France}
\affil[3]{Instituto de Estad\'{\i}stica (IESTA), Departamento de Métodos Cuantitativos, FCEA, Universidad de la Rep\'ublica, Montevideo, Uruguay}
\affil[4]{Computer, Electrical and Mathematical Sciences and Engineering Division (CEMSE), King Abdullah University of Science and Technology (KAUST), Saudi Arabia}
\affil[5]{Alexander von Humboldt Professor in Mathematics for Uncertainty Quantification, RWTH Aachen University, Germany}

\maketitle


\begin{abstract}

In this work, we extend the data-driven It\^{o} stochastic differential equation (SDE) framework for the pathwise assessment of  short-term forecast errors to account for the time-dependent upper bound that naturally constrains the observable historical data and forecast. We propose a new nonlinear and time-inhomogeneous SDE model with a Jacobi-type diffusion term for the observable phenomenon of interest, simultaneously driven by the forecast and the constraining upper bound.
We rigorously demonstrate the existence and uniqueness of a strong solution to the SDE model by imposing a condition for the time-varying mean-reversion parameter appearing in the drift term.  After normalization, the original forecast function is thresholded to keep such time-varying mean-reversion parameters bounded. Thus, for any finite time interval, the paths of the forecast error process almost surely do not reach the time-dependent boundaries.
The SDE model parameter calibration procedure is applied to user-selected approximations of the likelihood function. Another novel contribution is estimating the unknown transition density of the forecast error process with a tailored kernel smoothing technique with the control variate method, coupling an adequate SDE to the original one. We provide the theoretical study about how to choose the optimal bandwidth.
As a case study, we fit the model to the 2019 photovoltaic (PV) solar power daily production and forecast data in Uruguay, computing the daily maximum solar PV production estimation. Two statistical versions of the constrained SDE model are fit, with the beta and truncated normal distributions as surrogates for the transition density function of the forecast error process.
Empirical results include simulations of the normalized solar PV power production and pathwise confidence bands generated with the desired coverage probability through an indirect inference method. An objective comparison of optimal parametric points associated with the two selected statistical approximations is provided by applying our innovative kernel smoothing estimation technique of the transition function of the forecast error process. As a byproduct, we created a procedure providing a reliable criterion for choosing an adequate density proxy candidate that better fits the actual data at a low time cost. The methodology employs a thorough pathwise assessment of the forecast error uncertainty in situations where time-dependent boundaries and available forecasts drive the model specifications.

\vspace{0.3cm}

\vspace{0.3cm}
\end{abstract}



\section{Introduction}  \label{Section_1}

Assessing the time evolution of the forecast error is crucial in many disciplines, providing a global view of the mutual relationships between the observed phenomenon and its forecast. Often, measurable phenomena take values between time-dependent boundaries estimable using physics-based information or other sources.
The probabilistic modeling of the forecast error based on data-driven It\^{o} stochastic differential equations (SDEs) is a recent approach that constitutes a flexible framework for the user to incorporate such external inputs. This paper extends our previous work (\hcite{2021ckst}), including a time-dependent upper bound in SDE modeling. The proposed approach is fully illustrated by providing a case study based on solar photovoltaic (PV) power production data in Uruguay during 2019, its forecast, and the estimated maximum solar PV power production (the time-dependent upper bound in this case) built on a form of the ``very clear sky model'' (\hcite{sandiaGHI}, pp.~14--15).

As highlighted in the review (\hcite{snbmbm}, Section 4), SDEs have been developed in several recent studies on wind and solar power forecasting, generally expressed as state-space models for model estimation with noisy data. In work by \hciteauthor{immm}, three SDE models for solar irradiance were introduced, the first with a drift term designed to revert to a locally scaled numerical weather prediction provided by the Danish Meteorological Institute, with a constant reversion parameter $\theta_X$. In the other two models, the authors changed the diffusion coefficient from a constant parameter $\sigma_X$ to the nonlinear state-dependent term $\sigma_X X_t (1 -X_t)$. They also increased the structure of the drift term to account for some physical characteristics of the solar irradiance in the second model, in particular, replacing the constant reversion parameter $\theta_X$ with a stochastic process, and reverting to a fixed value in the third model.
In work by \hciteauthor{bggk}, the Clear Sky Index is modeled, defined as the ratio between the observed and maximal theoretical irradiance corresponding to an uncloudy sky, assuming perfect data observation. The authors proposed a SDE model with mean-reversion drift term to the deterministic forecast with constant reversion parameter, and a generalized diffusion coefficient expressed as $ \sigma X_t^{\beta} (1 - X_t)^{\gamma}$, $\frac{1}{2} \leq \beta, \gamma$.

Unlike the previous work, we specify the drift coefficient of the SDE model as in our preceding paper (\hcite{2021ckst}), using a mean-reverting term with a time-varying speed of reversion while tracking the time derivative of the deterministic forecast. In addition, a novel feature is that the diffusion coefficient of the SDE model for the normalized phenomenon of interest includes a deterministic time-dependent upper bound $h_t$ for the observed phenomenon and its forecast in the form of $\sqrt{2 \alpha \theta_0 X_t (h_t - X_t)}$, where $\theta_0 > 0$ is the baseline speed for mean-reversion, and $\alpha >0$ controls the path variability.
As we broadly reveal in this paper, such a model with a time-dependent upper bound function is suitable for assessing the uncertainty of the forecast error process of solar PV power production.
In our previous paper (\hcite{2021ckst}), SDE models were applied to assess the forecast error of normalized wind power production. This approach extends the SDE formulation presented in previous work (\hcite{mozuma}, \hcite{elkant}, \hcite{elkate}) by considering the time derivative tracking of the forecast in the drift term in the same spirit as in work by \hciteauthor{immm2} for wind speed modeling to avoid fitting a model with a systematic lag behind the forecast.
\hciteauthor{immtm} revealed how to include the wind speed-to-wind power relationship in a simplified version of the SDE system proposed by \hciteauthor{immm2} for noisy data by introducing a dynamic power curve in the SDE. From the perspective of the probabilistic forecast evaluation, \hciteauthor{bmm} illustrated how to apply scoring rules on a bounded-point forecast-driven SDE model and how to simulate six-dimensional predictive distributions from the SDE model considered by \hciteauthor{mozuma}, whose diffusion coefficient after the Lamperti transformation is independent of the state.

A primary contribution of this paper is demonstrating how to formulate a parsimonious parametric SDE model and its constraining time-dependent upper bound to develop a rigorous and versatile approach for probabilistic modeling of the forecast error. We highlight that the drift term of the SDE model contains a mean-reversion time-varying parameter whose boundedness is ensured by keeping the normalized forecast function away from the boundaries through a threshold parameter. The exact transition density of the proposed SDE model is unavailable in a closed analytic form. Given the statistical approximation specified by the user of the transition density of the forecast error process, we provide an optimization procedure that achieves the automatic calibration of the threshold parameter and the other SDE model parameters. 

Another innovative contribution is a nonparametric statistical algorithm to obtain the point estimation of the transition density of the forecast error process by applying a kernel smoothing technique with the control variate method and coupling an adequate SDE to the original one. Using this algorithm, we estimate the likelihood value of the SDE model. We develop a theoretical analysis for choosing the optimal bandwidth. In addition, our approach allows for checking, on the log-likelihood scale, the performance of the calibration procedure at the optimal parametric points derived employing the user-selected statistical approximations of the SDE model.

The rest of the paper is organized as follows. Section~\ref{Section_2} introduces the proposed framework based on data-driven constrained SDEs, stating the main theoretical results characterizing the class of  It\^{o} processes useful for analyzing forecast error uncertainty when the observable phenomenon and its forecast take nonnegative values below a time-varying upper bound. 
Section~\ref{Section_3} provides a statistical approach for calibrating the constrained SDE model based on the user choice of a surrogate distribution for the transition density of the forecast error process. The moment-matching technique is coupled with a fast ad hoc procedure to estimate the thresholding parameter of the normalized forecast function, presented in Appendix~\ref{app_varepsilon}, which is used as input to update the entire sample of forecast errors and optimize the model likelihood to obtain the diffusion parameter estimates. The likelihood value of the constrained SDE model is computed using a new nonparametric point estimator for the transition density of the forecast error process with the control variate method. The mathematical details to obtain an upper bound for the mean squared error (MSE) of the proposed nonparametric estimator are presented in Appendix~\ref{app_varbound}. 
Next, in Section~\ref{Section_4}, the methodology is applied to a data set with the normalized solar PV power production and its forecast in Uruguay during 2019.
The maximum solar PV power production is obtained through a simple irradiance clear sky solar model, whose characteristics are recalled in Appendix~\ref{app_irr}.
Final considerations and the outlook for future work are presented in Section~\ref{Section_5}. Appendix~\ref{app_datasets} displays further graphical data analysis results.



\section{Constrained stochastic differential equation model for the forecast error} \label{Section_2}

This section introduces a class of It\^{o} SDEs designed to model normalized observable phenomena driven by forecasts constrained below a time-varying upper bound.
The time-varying upper bound is a novel feature that generalizes the class of data-driven models introduced by \hciteauthor{2021ckst}, expanding their flexibility and the range of their applicability in real-world problems.
Moreover, we study the conditions under which the process, rescaled with the time-varying upper bound, is defined as a unique, strong solution to the SDE. The paths of the rescaled SDE process are subjected to a bounded reversion speed by truncating the forecast function with a threshold parameter. Finally, we express the rescaled forecast error process as the solution for an It\^{o} SDE model whose inference is developed in Section~\ref{Section_3}. \\

We let $(h_t)_{t \in [0,T]}$ be the computable upper bound function for the normalized observable phenomenon of interest. This function represents technical information we may include in the analysis. The solar PV power generation case is a practical example, where it is natural to assume that the maximum power generation varies over time. Indeed, this case is fully discussed in Section~\ref{Section_4}, presenting a real-world case study. \\ 

We assume that the temporal dynamic of the phenomenon $X = \{X_t \in [0, h_t], \, t \in [0, T] \}$ is modeled as the solution to the  It\^{o} SDE: 
\begin{equation}
\left\{\begin{array}{l}
d X_t=a(X_t ; p_t, \dot{p}_t, \boldsymbol{\theta}) dt + b(X_t ; h_t, \boldsymbol{\theta}) d W_t, \quad t \in[0, T] \\
X_0=x_0 \in[0,h_0]
\end{array}\right.
\label{eq:sde-general-solar}
\end{equation}

where

\begin{itemize}
\item $a(\cdot; p_t, \dot{p}_t, \bm{\theta}): [0, h_t] \to \mathbb{R}$ denotes a drift function,
\item $b(\cdot; h_t, \bm{\theta}): [0, h_t] \to \mathbb{R}^+$  represents a diffusion function,
\item $\bm{\theta}$ denotes a vector of unknown parameters,
\item $(p_t)_{t \in [0,T]}$ indicates a time-dependent deterministic function $[0, h_t]$-value that represents the available forecast, and $ (\dot{p}_t)_{t \in [0,T]}$ represents its time derivative,
\item $(h_t)_{t \in [0,T]}$ denotes a time-dependent deterministic function that represents the computable upper bound for the observed phenomenon and its forecast, and
\item $\{W_t, \, t \in [0,T] \}$ represents a standard real-valued Wiener process.
\end{itemize}

The time-dependent drift function for the process $X$ is initially specified as follows:
\begin{equation}
a(X_t; p_t, \dot{p}_t, \bm{\theta}) = \dot{p}_t  - \theta_t (X_t - p_t),  \label{drift:meanrev-derivtrack}
\end{equation} 
where $ (\theta_t)_{t \in [0,T]} $ is a positive deterministic function that depends on $\bm{\theta}$, the forecast function $p_t$, and its derivative $\dot{p}_t$. As indicated later in this section, $\theta_t$, the upper bound function $h_t$, and its derivative $\dot{h}_t$ satisfy a condition essential to guarantee that the nonnegative paths of $X$ are almost surely bounded above by the function $h$ at any finite period. 

As observed in work by \hciteauthor{2021ckst}, the drift specification (\ref{drift:meanrev-derivtrack}) enables the process $X$ to have the mean-reversion property with a time-varying speed $ \theta_t$, and it tracks the derivative term $ \dot{p}_t$. 
Applying It\^{o}'s lemma on $g(X_t, t) = X_t e^{\int_{0}^{t} \theta_s \dif s}$ immediately reveals that, given $  \mathbb{E}\left[X_0\right] = p_0$, the normalized process $X$ solving (\ref{eq:sde-general-solar}) is unbiased with respect to the forecast $p_t$ (i.e., $\mathbb{E} \left[X_t\right] = p_t$). \\

To ensure that the state space of $X_t$ is $[0, h_t]$ for every $t \in [0,T]$, we must select a convenient diffusion term and require that the time-varying parameter $ \theta_t$ satisfies an ad hoc condition. The state and time-dependent diffusion function is selected to prevent the process from exiting its natural range $[0,h_t]$ as follows:
\begin{equation}
    b (X_t; h_t, \bm{\theta} )= \sqrt{2 \alpha \theta_0 X_t (h_t - X_t)} \,,
  \end{equation}
where $\bm{\theta} = (\theta_0,\alpha)$, $\theta_0 > 0$ denotes an unknown baseline speed of mean-reversion, and $\alpha >0$ represents an unknown parameter that controls the path variability. 

At this stage, the proposed model is formulated as follows:
\begin{equation}
\left\{\begin{array}{l}
d X_t= \left(\dot{p}_t  - \theta_t (X_t - p_t) \right) d t +\sqrt{2 \alpha \theta_0 X_t (h_t - X_t)} d W_t, \quad t \in [0,T]  \\
X_0=x_0 \in[0,h_0]
\end{array}\right.
\label{ourmodel}
\end{equation}   
where $p_t$ indicates the normalized forecast, the time-varying parameter $\theta_t$ determines the strength of the mean-reverting effect, and $\alpha >0$ and $\theta_0 >0$ denote two unknown model parameters. 

We determine the conditions under which the nonstationary, and time-inhomogeneous process $X_t$ defined by (\ref{ourmodel}) exists and is unique. It is convenient to focus the analysis on the process $Y_t = \frac{X_t}{h_t}$ obtained through a change in the variable, which is well-defined for every $t \in [\tilde{t}, \tilde{T}]$, such that the $C^1$ function $h_t$ is strictly positive. \\ \noindent
The time points $\tilde{t} = \min\{ t> 0 : h_t >0 \}$ and $\tilde{T} = \max\{ t < T : h_t >0 \}$ are unique. 
It\^{o}'s formula leads to the following equation: 
\begin{equation}
\left\{\begin{array}{l}
d Y_t= \left(\dot{\left( \frac{p}{h} \right)}_{\!t}  - \left( \theta_t +  \frac{\dot{h}_t}{h_t} \right) (Y_t - \frac{p_t}{h_t}) \right) d t +\sqrt{2 \alpha \theta_0 Y_t (1 - Y_t)} d W_t, \quad t \in [\tilde{t}, \tilde{T}]  \\
Y_{\tilde{t}}=y_{\tilde{t}} \in[0, 1]\,,
\end{array}\right.
\label{ourmodel2}
\end{equation} 
where $\dot{\left( \frac{p}{h} \right)}_{\!t} \equiv \frac{d}{dt}  \!\left( \frac{p_t}{h_t}\right) = \frac{\dot{p}_t h_t - \dot{h}_t p_t}{h_t^2}$. \\

Remark: The SDE model (\ref{ourmodel2}) for the process $Y_t$ belongs to the family of data-driven diffusions that we introduced (\hcite{2021ckst}, p.~4), namely: 
\begin{equation*}
  \left\{
  \begin{array}{@{}rl@{}}
    dX_t \!\!\!&= (\dot{p}_t  - \theta_t (X_t - p_t) ) dt +\sqrt{2 \alpha \theta_0 X_t (1 - X_t)} dW_t\,, \;\; t \in [0,T]  \\
   X_0  \!\!\!&=  x_0 \in [0,1] \,,
 \end{array}\right. 
\end{equation*} 
In the present formulation (\ref{ourmodel2}), the role of $p_t$ is played by $\frac{p_t}{h_t}$, the time-varying parameter $\theta_t$ that rules the strength reversion changes to $\left(\theta_t +  \frac{\dot{h}_t}{h_t} \right)$. The derivative tracking term is updated from $\dot{p}_t$ to $\dot{\left( \frac{p}{h} \right)}_t$. \\ \noindent
We emphasize that the SDE model (\ref{ourmodel2}) for the process $Y_t$ includes the time-varying upper bound information $h_t$.

Consequently, the proof of the existence and uniqueness of the strong solution of process $Y_t$ to the SDE (\ref{ourmodel2}) for all $t \in  [\tilde{t}, \tilde{T}]$ with state space $[0, 1]$ is obtained by applying Theorem~1 from our previous paper (\hcite{2021ckst}, pp.~17--18), replacing $p_t$ with $\frac{p_t}{h_t}$, and $\theta_t$ with  $\left(\theta_t +  \frac{\dot{h}_t}{h_t} \right)$ (see Condition~\eqref{Assumption:1} in the theorem below). \\

We let $\tau_0:=\inf \{t \in [\tilde{t}, \tilde{T}],\; Y_t=0\}$ and  $\tau_1:=\inf \{t\in [\tilde{t}, \tilde{T}],\; Y_t=1\}$ with the convention that $\inf\emptyset=+\infty$. The following theorem summarizes the essential properties of the process $Y_t$, including the condition that allows the process $Y_t$ to avoid almost surely hitting the boundaries of the open interval $(0,1)$, whose proof applies Theorem~2 by \hcite{2021ckst}, p.~18.

\begin{Thm}\label{thm:exunasbo}
We assume that    
\begin{equation}\label{Assumption:1}
\forall  t \in  [\tilde{t}, \tilde{T}],\;\; 0\le \dot{\left( \frac{p}{h} \right)}_{\!t}  + \left( \theta_t +  \frac{\dot{h}_t}{h_t} \right)  \frac{p_t}{h_t} \le \theta_t +  \frac{\dot{h}_t}{h_t}  \le C_1 < \infty \,. \tag{A}    
\end{equation}
Then, a unique strong solution to (\ref{ourmodel2}) exists such that, for all $t \in  [\tilde{t}, \tilde{T}]$, $Y_t\in[0,1]\,.$

Moreover, if we assume that $y_{\tilde{t}} \in (0,1)$, $ \frac{p_t}{h_t}  \in (0,1)$  for all $t \in [\tilde{t}, \tilde{T}]$, and that Condition~\eqref{Assumption:1} is strengthened as follows 
\begin{equation}\label{Assumption:2}
 0 < \alpha \theta_0 \le \dot{\left( \frac{p}{h} \right)}_{\!t}  + \left( \theta_t +  \frac{\dot{h}_t}{h_t} \right)  \frac{p_t}{h_t} \le  \left( \theta_t +  \frac{\dot{h}_t}{h_t} \right) - \alpha \theta_0  \le C_1 < \infty \,, \tag{A$^\prime$}
\end{equation}
 then, $\tau_0=\tau_1=+\infty$ a.s.
\end{Thm}

Remark: Given $y_{\tilde{t}} \in (0, 1)$, the unique strong solution of (\ref{ourmodel2}) is guaranteed to have paths almost surely belonging to $(0,1)$ if the mean-reversion time-varying parameter $\left( \theta_t + \frac{\dot{h}_t}{h_t} \right)$ for the process $Y_t$ satisfies the following condition:
\begin{equation} \label{Assumption:B}
\theta_t + \frac{\dot{h}_t}{h_t} \geq \max \left(\frac{\alpha\theta_0+ \dot{\left( \frac{p}{h} \right)}_{\!t}}{1 - \frac{p_t}{h_t}},\frac{\alpha\theta_0 - \dot{\left( \frac{p}{h} \right)}_{\!t}}{\frac{p_t}{h_t}}\right)\,,  \:\: \forall t \in  [\tilde{t}, \tilde{T}] \,.\tag{B} 
\end{equation} 

Remark: Condition~\eqref{Assumption:B} reveals that, given a smooth upper bound $h_t$ and a smooth function $p_t$, it is generally impossible to determine a bounded time-varying parameter $\theta_t$ because the lower bound may become unbounded when $\frac{p_t}{h_t} \rightarrow 0$ or $\frac{p_t}{h_t} \rightarrow 1$. However, we consider the following thresholded normalized forecast function
\begin{equation}
\left(\frac{p_t}{h_t}\right)_{\!\!\varepsilon}=\begin{cases}
\varepsilon&\quad\text{if}\quad \frac{p_t}{h_t}<\varepsilon\\
\frac{p_t}{h_t}&\quad\text{if}\quad\varepsilon\leq \frac{p_t}{h_t}<1-\varepsilon\\
1-\varepsilon&\quad\text{if}\quad \frac{p_t}{h_t}\geq1-\varepsilon\,,
\end{cases}
\label{corrforecast}
\end{equation}
that satisfies $\left(\frac{p_t}{h_t}\right)_{\!\!\varepsilon} \in [\varepsilon, 1 - \varepsilon]$ for any $0 < \varepsilon  \ll \frac{1}{2}$ and $t \in [\tilde{t}, \tilde{T}]$, providing that $\theta_t$ is bounded for every $t \in [\tilde{t}, \tilde{T}]$.

In practice, tuning the threshold parameter $\varepsilon$ is arduous. In this paper, we calibrate $\varepsilon$ by using a fast ad hoc iterative procedure, that may retain most of the available information close to the boundaries, whose description is presented in  Appendix~\ref{app_varepsilon}. \\

Remark: Using Condition~\eqref{Assumption:B} with $\left(\frac{p_t}{h_t}\right)_{\!\!\varepsilon}$, a choice of the mean-reversion time-varying parameter $\theta_t$ 
that guarantees the existence and uniqueness of the strong solution of (\ref{ourmodel2}) in the interval $(0,1)$ is as follows: 

\begin{equation}
\theta_t=\max\left(\theta_0, \left[ \frac{\alpha\theta_0 + \big| \frac{d}{dt}  \!\left( \frac{p_t}{h_t}\right)_{\!\!\varepsilon} \big| }{\min \left( \left( \frac{p_t}{h_t}\right)_{\!\!\varepsilon}, 1 - \left(\frac{p_t}{h_t}\right)_{\!\!\varepsilon} \right)}  \right]   -\frac{\dot{h}_t}{h_t} \right)  \,,  \:\: \forall t \in  [\tilde{t}, \tilde{T}]\,.
\end{equation}

Finally, the change in variables $V_t = Y_t - \frac{p_t}{h_t} $ in (\ref{ourmodel2}), applied with the thresholded normalized forecast function (\ref{corrforecast}), leads to the following model for the forecast error process $\{V_t, t \in [\tilde{t}, \tilde{T}] \}$ of the rescaled observed phenomenon $Y_t  \equiv \frac{X_t}{h_t}$: 

\begin{equation}
\left\{\begin{array}{l}
d V_t=  - \left( \theta_t +  \frac{\dot{h}_t}{h_t} \right) V_t\, d t + \sqrt{2 \alpha \theta_0 \left(V_t + \left(\frac{p_t}{h_t}\right)_{\!\!\varepsilon}  \right) \left(1 - V_t - \left( \frac{p_t}{h_t}\right)_{\!\!\varepsilon} \right)} d W_t, \quad t \in [\tilde{t}, \tilde{T}]  \\
V_{\tilde{t}}=v_{\tilde{t}} \in \left[- \left( \frac{p_{\tilde{t}}}{h_{\tilde{t}}}\right)_{\!\!\varepsilon}, 1 - \left(  \frac{p_{\tilde{t}}}{h_{\tilde{t}}} \right)_{\!\!\varepsilon} \right],
\end{array}\right.
\label{VtSDE}
\end{equation}
\begin{itemize}
\item with  drift function $a(\cdot; p_t, \dot{p}_t, h_t, \dot{h}_t, \bm{\theta}, \varepsilon): \left[- \left( \frac{p_t}{h_t}\right)_{\!\!\varepsilon}, 1 - \left( \frac{p_t}{h_t}\right)_{\!\!\varepsilon} \right] \to \mathbb{R} $, and 
\item diffusion function $b(\cdot; p_t, h_t, \bm{\theta}, \varepsilon): \left[- \left( \frac{p_t}{h_t}\right)_{\!\!\varepsilon}, 1 - \left( \frac{p_t}{h_t}\right)_{\!\!\varepsilon} \right]  \to \mathbb{R}^+ $.
\end{itemize}

From now on, we retain the notation $\frac{p_t}{h_t}$ to denote the thresholded normalized forecast function (\ref{corrforecast}), unless specified otherwise. The next following section presents a complete inferential strategy for calibrating the threshold parameter $\varepsilon$ and diffusion parameters $ \bm{\theta} = (\theta_0, \alpha)$ in the constrained model (\ref{VtSDE}).


\section{Likelihood-based model inference and nonparametric estimation of the transition density of the forecast error process $V$} \label{Section_3} 

We developed a novel inferential methodology for the time-inhomogeneous nonlinear model (\ref{VtSDE}). The threshold parameter $\epsilon$, which intervenes in the It\^{o} SDE model  (\ref{VtSDE}) only through the data-driven part $\frac{p_t}{h_t}$, is calibrated through a fast ad hoc procedure presented in Appendix~\ref{app_varepsilon}. \\ \noindent
First, under the assumption of observing $M$ independent realizations of the process $V$ modeled by (\ref{VtSDE}), we expressed the likelihood function for the entire sample of forecast errors. Second, the transition densities of It\^{o}'s diffusion $V$ (the solution to (8)) characterized by the corresponding Fokker--Planck--Kolmogorov equation are not known analytically in closed form. Therefore, to make inference fast, we propose approximating the unknown densities adequately in the $V$-space, considering the constrained range of the paths of the process. After selecting a surrogate transition density, we used the moment-matching technique to obtain the expression for the approximated log-likelihood function.

\subsection{Likelihood function in the $V$-space}

Suppose $M$ non-overlapping paths of the continuous-time It\^{o} process $V = \{ V_t,  t \in [\tilde{t}, \tilde{T}]  \}$ are observed, each at $N_j + 1$ time instants, where $j=1, \ldots, M$, with sample frequency $\Delta$. We let $ V^{M,N}=\left\{ V_{{\tilde{t}_1}}^{N_1 + 1} , V_{{\tilde{t}_2}}^{N_2 + 1} ,\ldots , V_{{\tilde{t}_M}}^{N_M + 1} \right\}$ denote the entire random sample of forecast errors, where $V_{{\tilde{t}_j}}^{N_j + 1} =\left\{ V_{\tilde{t}_j + i \Delta}\,, i = 0, \ldots, N_j \right\}$, and $N \!=\! \sum_{j=1}^M (N_j+1)$. In addition, we let $\rho(v \vert v_{j, i-1}; p_{[t_{j,  i-1}, t_{j , i} ]}, h_{[t_{j,  i-1}, t_{j , i} ]}, \bm{\theta}, \varepsilon)$ denote the conditional probability density function of $V_{\tilde{t}_{j} + i \Delta} \equiv V_{j, i}$ given 
$V_{\tilde{t}_{j} + (i-1) \Delta} \equiv V_{j, i-1} = v_{j, i-1}$ evaluated at $v$. In this case, $p_{[t_{j,  i-1}, t_{j , i} ]}$  and $h_{[t_{j,  i-1}, t_{j , i} ]}$ are the observed forecast and upper bound functions at $[t_{j,  i-1}, t_{j , i} ] \equiv [\tilde{t}_j + (i-1) \Delta, \tilde{t}_j + i \Delta] $, $j = 1, \ldots, M$, $\, i = 1, \ldots, N_j$. Finally, $\bm{\theta} = (\theta_0, \alpha)$ and $\varepsilon$ are the unknown model parameters and unknown threshold, respectively.  \\ \noindent
The It\^{o} process $V$ defined by the SDE (\ref{VtSDE}) is Markovian. Under the assumption of independence among the $M$ realizations of $V$, the likelihood function of the sample $V^{M,N}$ can be written as follows:  

\begin{equation}
\mathcal{L}\left(\bm{\theta} ; V^{M,N},  \varepsilon \right) = \prod\limits_{j=1}^M \left\{ \prod\limits_{i=1}^{N_j} \rho \left( {V_{j, i}| V_{j, i-1}} ; p_{[t_{j,  i-1}, t_{j , i} ]},   h_{[t_{j,  i-1}, t_{j , i} ]}, \bm{\theta}, \varepsilon \right)    \right\},
\label{likelihood}
\end{equation}
where $t_{j ,i} \equiv  \tilde{t}_j + i \Delta$ for any $j = 1, \ldots, M$ and $i = 0, \ldots, N_j$. \\

Remark: The assumption of using a sample of independent realizations of the forecast error process is to be guaranteed in any application based on the statistical model (\ref{likelihood}). Section~\ref{Subsection_4-3} describes the simple strategy used in the actual case study for renewable energy from solar PV plants in Uruguay presented in Section~\ref{Section_4}.

\noindent The computation of the maximum likelihood estimates of $\bm{\theta}$ in (\ref{likelihood}) requires evaluating the transition density of the process $V$, which is analytically intractable in a closed form. Parameter estimation problems involving time-inhomogeneous nonlinear SDE models can be analyzed using approximate likelihood methods, as presented e.g., by \hcite{eglix}, \hcite[Chapter 11]{saso}, and \hcite[Section 3]{chls}. \\
In the proposed approach, similar to (\hcite{saso}, Section 11.4), a surrogate transition density function of the process $V$ is proposed. Next, the moment-matching technique, described below, provides the numerical framework to link the parameters of the surrogate density to the conditional moments of $V$  in any period.

\subsection{Moment-matching technique and approximate likelihood function in the $V$-space} \label{moments_ODEs}

This technique consists of matching the conditional moments of the SDE model (\ref{VtSDE}) to the moments of the selected surrogate transition density for $V$. We observe that the first conditional moment of the SDE model (\ref{VtSDE}) is given by  
$m_1(t) \equiv \mathbb{E} \left[V_t \vert V_{t_{j,i-1}} = v_{t_{j,i-1}} \right] =  v_{t_{j,i-1}} e^{- \int_{t_{j,i-1}}^t (\theta_s + \dot{h}_s/h_s) ds}$, for any $t\in [t_{j,i-1}, t_{j, i})$, $j = 1, \ldots, M$ and $i = 1, \ldots, N_j\,.$  For $k \geq 2$, we let $m_k(t) \equiv \mathbb{E} \left[V_t^k \vert V_{t_{j,i-1}} = v_{t_{j,i-1}} \right]$. Moreover, applying It\^{o}'s lemma with $g(V_t) = V_t^k$, we obtain

\begin{align}
\frac{d m_k(t) }{d t} & = - k \left[ \left( \theta_t  + \frac{\dot{h}_t}{h_t} \right) + (k-1) \alpha \theta_0 ) \right] m_k(t)   \nonumber \\ 
& + k (k-1) \alpha \theta_0   \left( 1 - \frac{2 p_t}{h_t} \right) m_{k-1}(t)   \nonumber \\ 
& + k (k-1) \alpha \theta_0 \left( \frac{p_t}{h_t}  - \left(\frac{p_t}{h_t}\right)^{\!\!2} \right) m_{k-2}(t) ,
\end{align}
with initial conditions $m_k(t_{j,i-1})= v_{t_j, i-1}^k \,.$ \\
For any $t\in [t_{j,i-1}, t_{j, i})$, the first two conditional moments of $V$ $\left( m_1(t) \text{ and } m_2(t) \equiv \mathbb{E}\left[V_t^2 \vert V_{t_{j,i-1}} = v_{t_j, i-1} \right] \right)$ are computed by solving the following system of ordinary differential equations (ODEs):

\begin{equation}
	\begin{cases}
		\frac{d  m_1 (t)}{d t} &=  - m_1(t)\left(\theta_t+\frac{\dot{h}_t}{h_t}\right)   \\
		\frac{d  m_2 (t)}{d t} &=  -2 \left[ \left( \theta_t +\frac{\dot{h}_t}{h_t} \right) +   \alpha \theta_0 \right] m_2(t) + 		                  2 \alpha\theta_0 \left( 1 - \frac{2 p_t}{h_t} \right) m_1(t) + 
	       2 \alpha\theta_0 \left( \frac{p_t}{h_t}  - \left(\frac{p_t}{h_t}\right)^{\!\!2} \right)
	\end{cases}
	\label{Vtmom}
\end{equation}
with initial conditions $m_1(t_{j,i-1})= v_{t_j, i-1}$ and $m_2(t_{j,i-1})= v_{t_j, i-1}^2 \,.$ 

\subsubsection{Beta distribution as the surrogate transition density in the $V$-space}

As a practical application of the moment-matching technique, we propose a surrogate beta transition density with shape parameters $(\xi_1, \xi_2)$ for the process $V$.
By equating the first two central conditional moments of $V$ with the corresponding moments of the surrogate beta distribution on 
$[-1 + \varepsilon, 1 - \varepsilon]$, for any $t\in [t_{j,i-1}, t_{j, i})$, the shape parameters are given by
\begin{equation}
 \begin{split}
\xi_1(t) & = - \frac{(\mu(t) + 1 - \varepsilon)(\mu(t)^2 + \sigma(t)^2 - (1- \varepsilon)^2)}{2 (1 - \varepsilon) \sigma(t)^2}, \\ 
\xi_2(t) & =  \frac{(\mu(t) - 1 + \varepsilon )(\mu(t)^2 + \sigma(t)^2 - (1- \varepsilon)^2)}{2 (1 - \varepsilon) \sigma(t)^2} , \label{param_transformed_beta}
 \end{split}
 \end{equation}
where $\mu(t) = m_1 (t)$ and $\sigma(t)^2= m_2 (t)- m_1 (t)^2\,.$ 
The approximate log-likelihood $\tilde{\ell}(\cdot ; v^{M, N}, \varepsilon)$ of the observed sample $v^{M, N}$ can be expressed as
\begin{equation}
 \tilde{\ell} \left(\bm{\theta} ; v^{M,N},  \varepsilon \right) 
 = \sum_{j=1}^M \sum_{i=1}^{N_j} \log  \Bigg\{ \frac{1}{2 (1 - \varepsilon)} \frac{1}{B(\xi_1(t_{j,i}^-), \xi_2(t_{j,i}^-))} \left( \frac{v_{j,i} + 1 - \varepsilon}{2 (1 - \varepsilon)} \right)^{\xi_1(t_{j,i}^-) -1}  \left( \frac{ 1 - \varepsilon - v_{j,i}}{2 (1 - \varepsilon)} \right)^{\xi_2(t_{j,i}^-) -1} \Bigg\},
\label{eq:loglikelihoodV-beta}
\end{equation}
where the shape parameters $\xi_1(t_{j,i}^-)$ and $\xi_2(t_{j,i}^-)$, according to (\ref{param_transformed_beta}), depend on the limit quantities $\mu(t_{j,i}^-;\bm{\theta}, \varepsilon )$ and $\sigma^2(t_{j,i}^-;\bm{\theta}, \varepsilon )$ as $t\uparrow t_{j,i}$ computed by numerically solving the initial-value problem (\ref{Vtmom}). Further, $B(\xi_1,\xi_2)$ denotes the beta function. \\

Remark: The proposed approximation strategy for the likelihood function (\ref{likelihood}) is flexible and user-friendly. The user specifies the form of the surrogate transition density of $V$ suitable for the problem under study. Next, the numerical solution of a system of ODEs allows for characterizing the shape of the approximate log-likelihood given the available information. 

\subsubsection{Truncated normal distribution as the surrogate transition density in the $V$-space}

As another example, choosing a truncated normal distribution as a surrogate transition density for $V$, with support $[-1 + \varepsilon,1 - \varepsilon]$, leads to the following approximate log-likelihood $\tilde{\ell}(\cdot ; v^{M, N}, \varepsilon)$ of the observed sample $v^{M, N}$:
\begin{equation}
	\begin{aligned}
		\tilde{\ell} \left(\bm{\theta} ; v^{M,N}, \varepsilon \right) 
		&= \sum_{j=1}^M \sum_{i=1}^{N_j} \log \Bigg\{\frac{1}{\sigma(t_{j,i}^-)} \frac{\phi\left(\frac{x-\mu(t_{j,i}^-)}{\sigma(t_{j,i}^-)}\right)}{\Phi\left(\frac{1 - \varepsilon -\mu(t_{j,i}^-)}{\sigma(t_{j,i}^-)}\right)-\Phi\left(\frac{-1 + \varepsilon -\mu(t_{j,i}^-)}{\sigma(t_{j,i}^-)}\right)}
		\Bigg\},
		\label{eq:loglikelihoodV-truncated-gaussian}
	\end{aligned}
\end{equation}
where $\phi(\cdot)$ is the probability density function of the standard normal distribution, $\Phi(\cdot)$ is its cumulative distribution function and $\mu(t_{j, i}^-)$ and $\sigma(t_{j, i}^-)$ are the mean and standard deviation parameters, respectively.

\subsection{Calibration procedure for approximate log-likelihoods of constrained SDE models} \label{iter2}

Given an expression of the approximate likelihood function, adequate to the problem, we first use a fast ad hoc iterative procedure to recover the threshold parameter $\varepsilon$, presented in  Appendix~\ref{app_varepsilon}. 

Once the calibrated estimate ${\hat{\varepsilon}}$ of the threshold parameter is obtained, we apply it to the thresholded normalized forecast function $\left(\frac{p_t}{h_t}\right)_{\!\!\varepsilon}$ in (\ref{corrforecast}) deriving the adjusted observed forecast errors. Next, we compute the initial values of the diffusion parameters using formulas (\ref{initheta0}) and (\ref{initheta0alpha}), but for the entire data set, as done in (\hcite{2021ckst}, Subsection 4.4.1, pp.~7--8).

As the last step for model calibration, given the calibrated $\hat{\varepsilon}$ and initial guesses for $(\theta_0, \alpha)$, we minimize the negative log-likelihoods (\ref{eq:loglikelihoodV-beta}) and (\ref{eq:loglikelihoodV-truncated-gaussian}) over $(\theta_0, \alpha)$. Therefore, given any surrogate that the user may specify for the transition density of the process $V$, we obtain the optimal maximum likelihood estimates for the respective approximated statistical version of the model (\ref{VtSDE}). 

The following section introduces an innovative nonparametric procedure to compute the log-likelihood (\ref{likelihood}) of the SDE model (\ref{VtSDE}) at any previously derived optimal parametric point $({\hat{\varepsilon}}, {\hat{\theta}}_0, {\hat{\alpha}})$.

\subsection{Kernel density point estimation of the transition function of the process $V$} 
\label{nonparKDE}
For any fixed parametric point $(\varepsilon, \theta_0, \alpha)$, we propose estimating the transition density of the forecast error process $V$ by applying an innovative kernel smoothing technique with the control variate method. Therefore, we may evaluate the entire log-likelihood function of the SDE model (\ref{VtSDE}), enabling an objective comparison between the optimal parametric points obtained by employing statistical versions of (\ref{VtSDE}) selected by the user. In the case study presented in Section~\ref{Section_4}, we assess the performance between the statistical versions given by the beta and truncated normal distributions as surrogates of the transition density of $V$. \\

To start, we consider the system of two coupled SDEs, for any $ t \in [\tilde{t}, \tilde{T}]$, driven by the same Wiener process $W_t$, given by model (\ref{VtSDE}) and the following model for the auxiliary process $Z_t$: 
\begin{equation}
		\left\{\begin{array}{l}
			d Z_t=  - \left( \theta_t +  \frac{\dot{h}_t}{h_t} \right) Z_t\, d t + \sigma_t d W_t, \quad t \in [\tilde{t}, \tilde{T}]  \\
			Z_{\tilde{t}}=z_{\tilde{t}} \in \mathbb{R}\,,
		\end{array}\right.
		\label{ZtSDE}
	\end{equation}   
where $z_{\tilde{t}} = v_{\tilde{t}}$. The coordinates of the parametric point  $(\theta_0, \alpha, \varepsilon)$ are known values. The drift function of the model (\ref{ZtSDE}) has the same structure as the model (\ref{VtSDE}). Instead, the diffusion function of the model  (\ref{ZtSDE}), $\sigma_t$, is to be estimated. For this purpose, we obtained $\sigma_{j, i}$ using a moment-matching procedure, by setting the first two moments of (\ref{VtSDE}) and (\ref{ZtSDE}) to be equal. For any 
$t \in [t_{j,  i-1}, t_{j , i} ) \equiv [\tilde{t}_j + (i-1) \Delta, \tilde{t}_j + i \Delta)$, $j = 1, \ldots, M$, $\, i = 1, \ldots, N_j$, with $\tilde{t}_j$ set as the starting time of the $j$th path $V_{{\tilde{t}_j}}^{N_j + 1} =\left\{ V_{\tilde{t}_j + i \Delta}\,, i = 0, \ldots, N_j \right\}$ of the process $V$, the moments of $Z$ are derived by solving the following system of ODEs:
\begin{equation}
		\begin{cases}
			\frac{d  m_1 (t)}{d t} &=  - m_1(t)\left(\theta_t+\frac{\dot{h}_t}{h_t}\right)   \\
			\frac{d  m_2 (t)}{d t} &=  -2  \left( \theta_t +\frac{\dot{h}_t}{h_t} \right) m_2(t)+\sigma_t^2\,,
		\end{cases}
		\label{Ztmom}
	\end{equation}
with the initial conditions $m_1(t_{j,i-1})= v_{t_j, i-1}$ and $m_2(t_{j,i-1})= v_{t_j, i-1}^2 \,.$ 

Using the Euler scheme (\hcite{iacus1}, p.~62), for every path $j$ and time instant $i$, we draw a sample of size $m$, \\ $\left\{ \left(V_{j,i}^{(n)},Z_{j,i}^{(n)}\right), n = 1,\dots, m \right\}$ of the coupled process $\left(V,Z\right)$. The proposed estimator of the transition density of the process $V$ at the observed value $v_{j,i}$ is a kernel density estimator with control variate given by
\begin{equation}
	{\widehat{\rho}}_m(v_{j,i} ; h) :=\frac{1}{m}\sum_{n=1}^{m}\left(\kappa_{h}(V^{(n)}_{j,i}-v_{j,i})-\kappa_{h}(Z^{(n)}_{j,i}-v_{j,i})\right) + \mathbb{E} \left[\kappa_{h}\left(Z_{j,i} - v_{j,i}\right)\right]\,,
	\label{kde}
\end{equation}
where $\kappa_{h}(\cdot) = \frac{1}{h} \kappa\left( \frac{\cdot}{h}\right)$ denotes, as usual, the rescaled kernel with bandwidth $h$, and $\kappa(\cdot)$ is set to $\phi(\cdot)$.  
To select the bandwidth $h$ of the estimator (\ref{kde}) in an optimal way, we propose using the error control given by Theorem \ref{thm:msekde} in Appendix~\ref{app_varbound} and then minimize an upper bound for the corresponding MSE.  Recall that $\left\{ V_{j,i}^{(n)}, n = 1,\dots, m \right\}$  are independent copies of $V$, solution to \eqref{VtSDE} which almost surely belongs to the compact set $\left[- \left( \frac{p_t}{h_t}\right)_{\!\!\varepsilon}, 1 - \left( \frac{p_t}{h_t}\right)_{\!\!\varepsilon} \right]$. 
This ensures the boundedness of the density  $\rho_V$ of $V$ on its support. However, proving  the smoothness of  $\rho_V$ is clearly out of the scope of this paper, since the diffusion coefficient of \eqref{kde} is not globally Lipschitz and both drift and diffusion coefficients are time-dependent.  Despite of this fact, the upper bounds given by Theorem \ref{thm:msekde} are still of practical interest. It follows that the MSE of  (\ref{kde}) satisfies 
$$\begin{aligned}
	\text{MSE}\left\{{\widehat{\rho}}_m(v_{j,i} ; h)\right\} \leq C_1 h^4+\frac{C_{2}}{m \,h^{\frac{7}{2}}}, \;\;
	\text{where} \;\;
	\left\{\begin{array}{l}
		C_{1}= \frac{\|\rho^{\prime\prime}_{V_{j,i}}\|^2_{\infty}}{4}\\
		C_{2}= \frac 1{\pi}{\mathbb E}^{\frac 12}\left[ \Bigl| (V_{j,i}-v_{j,i})^2 - (Z_{j,i}-v_{j,i})^2 \Bigr|^2 \right] 
\sqrt{ \left(\sqrt{\pi}\|\rho_{V_{j,i}}\|_{\infty}+ \frac{1}{\sqrt{2} \sigma_{Z_{j,i}}}\right)}
	\end{array}\right.
\end{aligned}$$  
with $ \sigma^2_{Z_{j,i}}:=\text{Var}[Z_{j,i}]$.
Thus, given a sample of size $m$, the optimal choice of the bandwidth $h$ is given by
\begin{equation}
	h^{*}(m) =\left(\frac{7C_2}{8 m C_1}\right)^{\frac{2}{15}}
	\label{h_opt}
\end{equation}
which leads to an optimal $\text{MSE}\left\{{\widehat{\rho}}_m(v_{j,i} ; h^*(m))\right\}=O(m^{-\frac 8{15}})$. However, if we optimize the classical density kernel estimator without control variate given by 
$$
{\tilde{\rho}}_m(v_{j,i} ; h) :=\frac{1}{m}\sum_{n=1}^{m}\kappa_{h}(V^{(n)}_{j,i}-v_{j,i}),
$$
we get 
\begin{align}\label{var:classic}
\mbox{Var}\left({\tilde{\rho}}_m(v_{j,i} ; h) \right)&\le \frac{1}{2\pi mh^2}\mathbb E\left[ \exp\left(-\frac{(V_{j,i}-v_{j,i})^2}{h^2}\right)\right]\notag\\
&= \frac{1}{2\pi mh}\int_{\mathbb R} \exp\left(-y^2\right)\rho_{V_{j,i}}(hy+v_{j,i})dy\le  \frac{\|\rho_{V_{j,i}}\|_{\infty}}{2\sqrt{\pi} mh}.
\end{align}
As the bias for both estimators ${\widehat{\rho}}_m(v_{j,i} ; h)$ and ${\tilde{\rho}}_m(v_{j,i} ; h)$ is the same, we deduce that 
$$
\text{MSE}\left\{{\widehat{\rho}}_m(v_{j,i} ; h)\right\} \leq {C}_1 h^4+\frac{{\tilde C}_{2}}{m \,h} \mbox{ with } {\tilde C}_2:=
\frac{\|\rho_{V_{j,i}}\|_{\infty}}{2\sqrt{\pi}}.
$$
An elementary optimization gives ${\tilde h}^*(m):=\left(\frac{{\tilde C}_2}{4 m { C}_1}\right)^{\frac{1}{5}}$ which leads to 
\begin{equation}\label{MSE:classic}
\text{MSE}\left\{{\tilde {\rho}}_m(v_{j,i} ; {\tilde h}^*(m))\right\}=O(m^{-\frac 4{5}})<<\text{MSE}\left\{{\widehat{\rho}}_m(v_{j,i} ; h^*(m))\right\}=O(m^{-\frac 8{15}}).
\end{equation}
These calculations suggest  to choose a different $h^*$ for our estimator ${\widehat{\rho}}_m(v_{j,i} ; h)$. According to \eqref{kde}, we can write
\begin{align*}
\mbox{Var}\left({\widehat{\rho}}_m(v_{j,i} ; h) \right)= \mbox{Var}\left({\tilde{\rho}}_m(v_{j,i} ; h)\right) +
\frac 1m \mbox{Var}\left(\kappa_{h}(Z^{(n)}_{j,i}-v_{j,i})\right)  - \frac 2m\mbox{Cov}\left(\kappa_{h}(V^{(n)}_{j,i}-v_{j,i}),\kappa_{h}(Z^{(n)}_{j,i}-v_{j,i})\right)>0.
\end{align*}
Clearly, the main interest in using the control variate $Z_{j,i}$ is to have a positive covariance term that allows variance reduction. Therefore, by the same arguments as in \eqref{var:classic} and \eqref{MSE:classic}, we get 
\begin{align*}
\mbox{Var}\left({\widehat{\rho}}_m(v_{j,i} ; h) \right)&\le  \mbox{Var}\left({\tilde{\rho}}_m(v_{j,i} ; h)\right) +
\frac 1m \mbox{Var}\left(\kappa_{h}(Z^{(n)}_{j,i}-v_{j,i})\right)\\
&\le   \frac{\|\rho_{V_{j,i}}\|_{\infty} +\frac1{\sqrt{2\pi}\sigma_{Z{j,i}}}}{2\sqrt{\pi} mh}:=\frac{\hat C_2}{mh}
\end{align*}
and consequently 
$$
\text{MSE}\left\{{\widehat{\rho}}_m(v_{j,i} ; h)\right\} \leq C_1 h^4+\frac{\hat C_{2}}{mh}.
$$
In this case, an elementary optimization leads to choose ${\hat h}^*(m):=\left(\frac{{\hat C}_2}{4 m {C}_1}\right)^{\frac{1}{5}}$
and then 
$$
\text{MSE}\left\{{\hat{\rho}}_m(v_{j,i} ; {\hat h}^*(m))\right\}=O(m^{-\frac 4{5}})<<\text{MSE}\left\{{\hat{\rho}}_m(v_{j,i} ; {h}^*(m))\right\}=O(m^{-\frac 8{15}})
$$
with taking advantage of a variance reduction effect thanks to the introduced control variate. \\

Remark: Theorem \ref{thm:msekde} can be easily generalized using other kernels and quadrature rules. \\
Finally, we define the relative MSE of ${\widehat{\rho}}_m(v_{j,i}; h)$ as follows:	
\begin{equation}
 MSE_{rel}\left\{{\widehat{\rho}}_m(v_{j,i}; h)\right\} = \frac{MSE\left\{{\widehat{\rho}}_m(v_{j,i}; h)\right\}}{1+\rho_{Z}( v_{j,i})^2}\,.
 \label{MSE_rel}
\end{equation}
If necessary, we double the sample size $m$ to compute the estimator (\ref{kde}), repeating the previous procedure until the relative MSE (\ref{MSE_rel}) meets the required accuracy. \\

The following pseudocode describes the entire algorithm for the nonparametric point estimation of the transition density of the forecast error $V$ using kernel smoothing with control variates. The pseudocode ensures an adaptive determination of the sample size of the estimator (\ref{kde}) until the relative MSE is less than or equal to the prescribed error tolerance.

\begin{algorithm}[H]
	\caption{Kernel density point estimation of the transition function of the process $V$} 
	\label{KDE}
	\begin{algorithmic}[1]
		\State \textbf {Input}: $\hat{\theta}_0$, $\hat{\alpha}$, $\hat{\varepsilon}$, $p=p_{j,i}$, $h=h_{j,i}$, $\dot{h}=\dot{h}_{j,i}$, $j \in \{1,\dots,M\}$, $i  \in  \{0,\dots, N_j \!-\!1\}$, time spacing $\Delta$, m=50
		\For{\texttt{$j=1 \textrm{ to } M$}}
		\State\text{Initialization}: Set $Z_{j,0}=Y_{j,0}$
		\For{\texttt{$i= 0 \textrm{ to }  N_j - 1$}}
		\State\text{Compute} $\sigma_{j,i} \gets$\textbf{matching} the moments of $Z$ with the moments of $V$
		\State\text{Generate} $\Delta W_{i} \sim \mathcal{N}(0,\Delta)$ 
		\For{\texttt{$n=1 \textrm{ to } m$}}
		\State $V_{j,i+1}^{(n)}=V_{j,i}^{(n)}-\left(\theta_i-\frac{\dot{h}_{j,i}}{h_{j,i}}\right) V_{j,i}^{(n)}\Delta+\sqrt{2\alpha\theta_{0}\left(V_{j,i}^{(n)}+\left( \frac{p_{j,i}}{h_{j,i}}\right)\right)\left(1-V_{j,i}^{(n)}-\left( \frac{p_{j,i}}{h_{j,i}}\right)\right)} \Delta W_{i}$
		\State $Z_{j,i+1}^{(n)}=Z_{j,i}^{(n)}-\left(\theta_i-\frac{\dot{h}_{j,i}}{h_{j,i}}\right)Z_{j,i}^{(n)}\Delta+\sigma_{j,i} \Delta W_{i}$
		\EndFor
		\State Compute $h^*$ from Eq.~(\ref{h_opt})
		\State Compute $MSE_{rel}$ from Eq.~(\ref{MSE_rel})
		\State Compute ${\widehat{\rho}}_m(v_{j,i} ; h^*)$ from Eq.~(\ref{kde})
		\If{\texttt{$MSE_{rel}>0.1$ or ${\widehat{\rho}}_m(v_{j,i} ; h^*)  <0 $}}
		\State $m=2\times m$
		\State Repeat computations of Lines 11, 12, and 13
		\EndIf
		\EndFor
		\EndFor
	\end{algorithmic} 
\end{algorithm}

The following section focuses on the presentation of an actual case study in the field of renewable energy, demonstrating the use of the two statistical approximations of the log-likelihood of the model~(\ref{VtSDE}) based on the beta (\ref{eq:loglikelihoodV-beta}) and the truncated normal (\ref{eq:loglikelihoodV-truncated-gaussian}) densities, introduced in Section~\ref{moments_ODEs}. The performance of the calibration procedure at the optimal parametric points for the two selected statistical approximations is assessed by computing the value of the logarithm of the likelihood (\ref{likelihood}) of the SDE model (\ref{VtSDE}) in such points. Such computations rely on the innovative kernel density estimation procedure for the pointwise evaluation of the transition density of the process $V$, described in 
Section~\ref{nonparKDE}.\\



\section{Application: 2019 daily Uruguay solar photovoltaic power production and forecast data set} \label{Section_4}

Daily solar PV power production and forecast observations in Uruguay during 2019 normalized for the 228.8~MW maximum installed solar power capacity in this year (for more details, see \hcite{Kha}) are publicly available from the National Administration of Power Plants and Electrical Transmissions (UTE), Uruguay's government-owned power company. 
Daily production recordings of the solar PV power are available every 10 min, and one-day-ahead forecasts are provided daily with hourly frequency.

\subsection{Computation of the daily maximum solar PV power production}

The PV power generated by solar panel plants through the direct conversion of sunlight to electricity depends upon meteorological and atmospheric conditions.
Under optimal conditions, assuming the total efficiency of the solar cells, the solar PV power cannot exceed a given maximum time-dependent power production. Thus, $h_t$ denotes such a maximum value at time $t$.
Every day of the year, it is essential to estimate the time-dependent upper bound for the solar PV power production $h_t$ because any realistic model for the normalized solar PV power production must take values, at any time $t$ between zero and the maximum given by $h_t$. This section focuses on estimating the time-dependent upper bound $h_t$, which is input, joint with its derivative $\dot{h}_t$, driving the SDE model (\ref{VtSDE}). \\

For any given day and location, we assume that the maximum solar PV power production can be expressed as a function of the irradiance, a measure of sunlight power in watts per square metre ($\text{W/m}^2$) in SI units. Several choices of irradiance are possible.
To provide a realistic upper bound of the solar PV system, we set $h_t \equiv h(t)$ equal to $ k \, I_{D}(t)$, where $I_{D}(\cdot)$ is the so-called direct normal irradiance, the radiation directly from the Sun measured as the flux of the beam radiation through a plane perpendicular to the direction of the Sun (\hcite{sandiaGHI}, p.~9; \hcite{Karathanasis2019}, p.~14). The term $k$ is an estimable constant measured in meters squared.

Many clear sky models with different degrees of complexity have been developed for estimating irradiance levels. 
We adopted a form of the Meinel model (1976) pertaining to the class of ``very clear sky model," where irradiance depends only on the elevation angle ($0^{\circ}$ at sunrise and $90^{\circ}$ when the Sun is directly overhead. It varies throughout the day, depending on the latitude of a particular location and the day of the year.) Specifically, the irradiance function is defined as
\begin{equation}
 I_{D}(\cdot)= \frac{ I_0 \times 0.7^{AM_r(\cdot)^{0.678}}}{1000}\,,  \label{eq:irrf}
\end{equation}
where $\frac{I_0}{1000}$ (in $\text{MW/m}^2$) denotes an estimate of the solar constant (\hcite{dbb20}, pp.~5--6; \hcite{SAYIGH1979}), and $AM_r(\cdot)$ is the empirical expression of the actual air mass that considers the refraction effect of the atmosphere on solar radiation propagation (\hcite{IQBAL198385}, p.~100). The details for the computation of the irradiance function (\ref{eq:irrf}) and the daily maximum solar PV power production $h(t)$ are presented in Appendix~\ref{app_irr}. The sign of the second derivative of the upper bound function $h(t)$, computed numerically for every day and time instant, is nonpositive at almost all points, indicating that its graph is concave down.

\begin{figure}[H]
\minipage{0.49\textwidth}
\includegraphics[width=\linewidth]{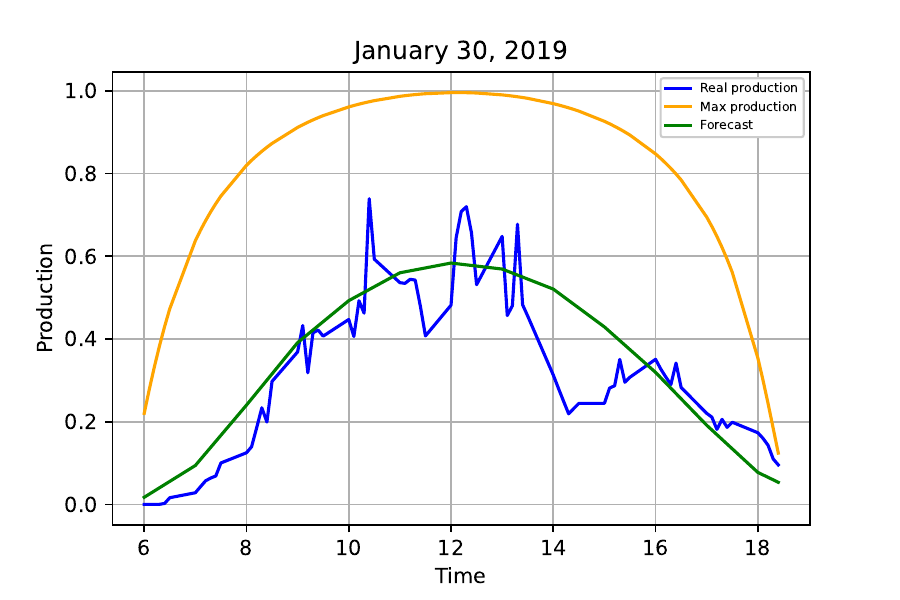}
\label{fig:example1}
\endminipage\hfill
\minipage{0.49\textwidth}
\includegraphics[width=\linewidth]{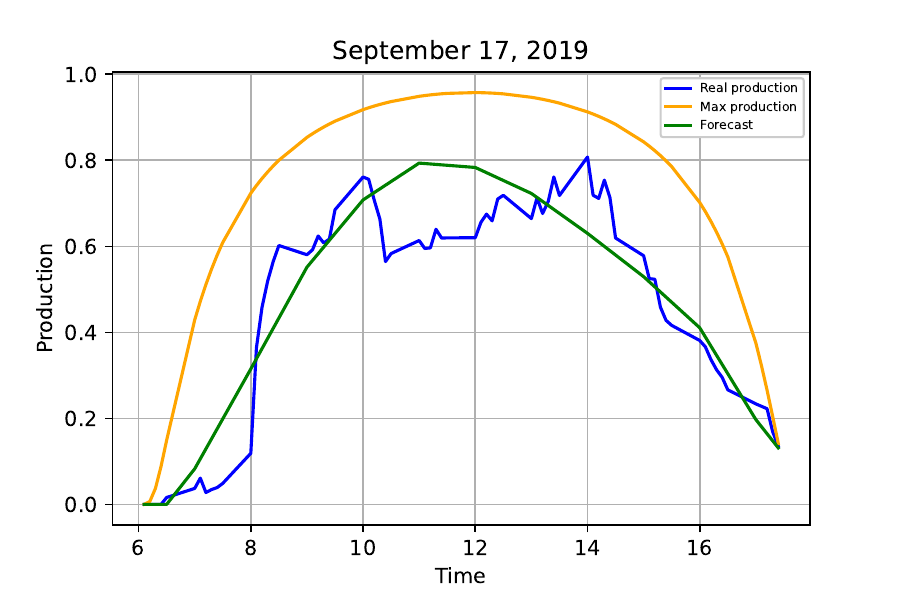}
\label{fig:example2}
\endminipage\hfill
\vspace*{-0.5cm}
\caption{Two daily segments with the normalized solar photovoltaic (PV) power production in Uruguay (blue line) recorded every 10 min, hourly solar PV power production forecast (green line), and upper bound (orange line).}
\label{fig:example-days}
\end{figure}

Figure~\ref{fig:example-days} presents two daily segments with their normalized upper bounds, real production, and one-day-ahead forecasted production adequately interpolated to equal the 10-min frequency of the real production.
\vspace{-0.7cm}

\subsection{Data summary}

A close inspection of the 2019 daily Uruguay solar PV power production and forecast data set reveals that several daily segments display signs of production curtailment due to the decision imposed by the electric system operators. 
After removing the days with signs of curtailment, the data set comprises 206 days (for more details, see work by \hciteauthor{Kha}). The accuracy of the forecast is seriously compromised by any curtailment decision taken posterior to the one-day-ahead forecasted production. 

Starting with the analysis of the forecast error of the normalized actual production $Y$, denoted by $V_t = Y_t - \frac{p_t}{h_t}$, according to the notation introduced in Section~\ref{Section_2}, we consider the mean absolute error (MAE) for every day and at every 10-min frequency with a strictly positive upper bound. We let $V_{j, i} \equiv V_{\tilde{t}_{j} + i \Delta}$ be the forecast error measured for the $j$th day, $j \in{\{1, \ldots, 206\}}$. The time instant is $i$, where $\tilde{t}_j$ is the starting time of the $j$th path $\left\{ V_{\tilde{t}_j + i \Delta}\,, i = 0, \ldots, N_j \right\}$ of the process $V$, $\Delta = 10 \text{[min]}$, such that $h_i > 0$.

The 10-min MAE $\left\{ \frac{1}{206} \sum_{j=1}^{206} \vert V_{j, i} \vert \,, \forall i : h_i >0 \right\}$ and the daily MAE  $\left\{ \frac{1}{\# \{i : h_i > 0 \}} \sum_{\{i : h_i > 0 \}} \vert V_{j, i} \vert \,, j = 1, \dots 206 \right\}$ are visualized below.

\begin{figure}[H]
\minipage{0.49\textwidth}
  \includegraphics[width=\linewidth]{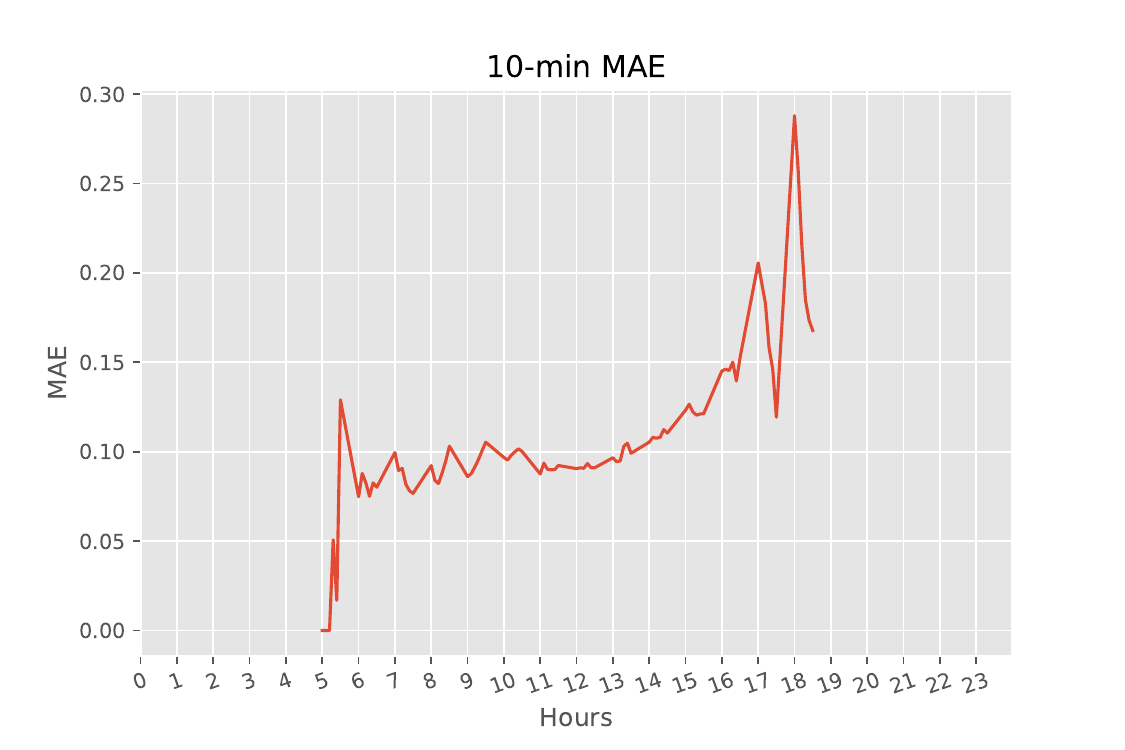}
  \caption{Ten-min mean absolute error (MAE).}\label{fig:hourly_mae}
\endminipage\hfill
\minipage{0.49\textwidth}
  \includegraphics[width=\linewidth]{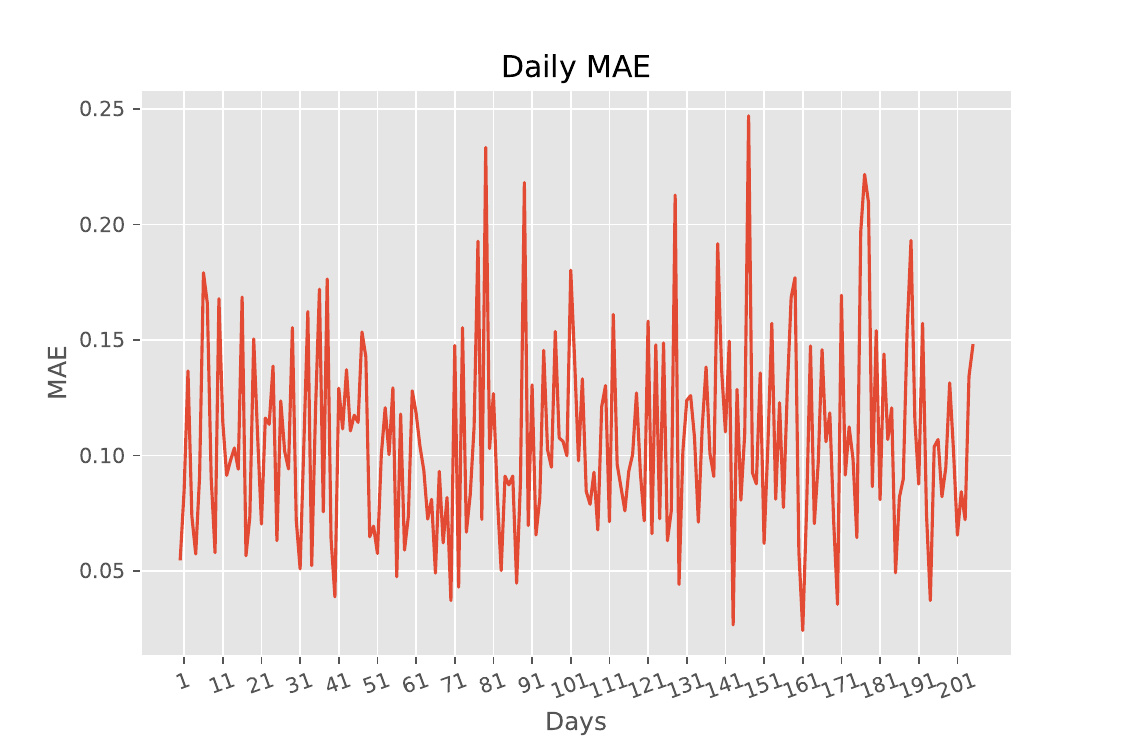}
  \caption{Daily  mean absolute error (MAE).}\label{fig:daily_mae}
\endminipage\hfill
\end{figure}

Figure~\ref{fig:hourly_mae} presents the 10-min MAE for the data set with 206 daily segments. The error becomes more pronounced in the afternoon, almost reaching 0.29 at 6 p.m.
Figure~\ref{fig:daily_mae} displays the pattern of the daily MAE for the data set with 206 daily segments.

Second, we analyze the forecast error transition density. The kernel density estimate based on the forecast error transitions for all solar PV power range, using a standard normal kernel and Scott's rule for bandwidth selection (\hcite{scott}, p.~164, eq.~6.44), is presented next. 

\begin{figure}[H]
    \centering
   \minipage{0.7\textwidth}%
  \includegraphics[width=\linewidth]{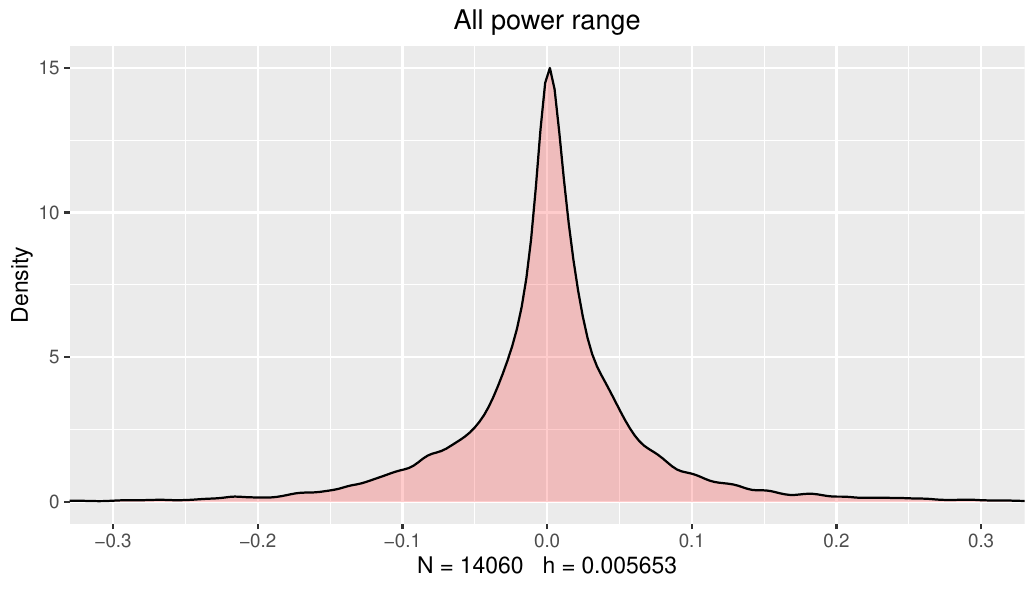}
  \caption{Kernel density estimation based on the solar photovoltaic power forecast error transitions for all data with standard normal kernel and Scott's bandwidth selection method.}
  \label{fig:error_forecast_all}
\endminipage
\end{figure}

Finally, Figure~\ref{fig:error_forecast_all} illustrates the scatterplot of the observed normalized forecast error $v_t$ versus the available normalized forecast $p_t$ for the data set with 206 daily segments.

\begin{figure}[H]
    \centering
   \minipage{0.7\textwidth}%
  \includegraphics[width=\linewidth]{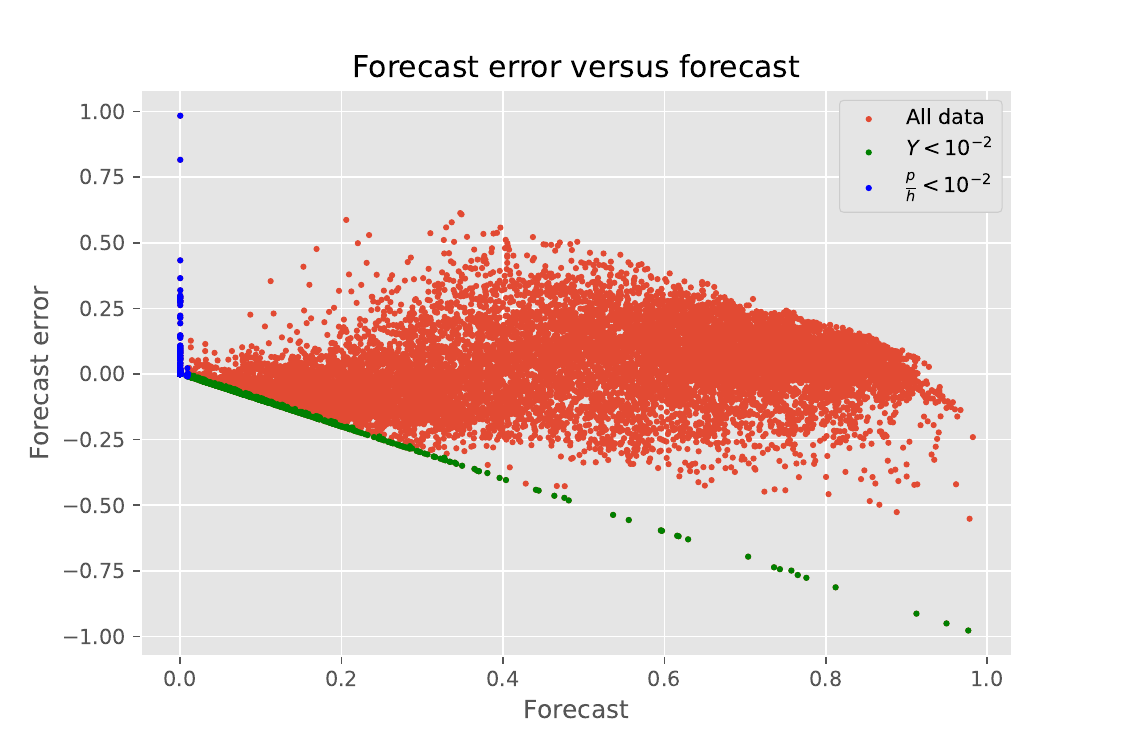}
  \caption{Forecast error $v_t$ versus forecast $p_t$.}
  \label{fig:error_forecast_all}
\endminipage
\end{figure}

\subsection{Application of the calibration procedure for approximate log-likelihoods} \label{Subsection_4-3}
We split the global 206 daily segments data set into two subsets, with 103 noncontiguous segments each, labeled Data Sets 1 and 2. All graphical data analyses presented for the entire data set in the previous section are reproduced separately for both data sets in Appendix~\ref{app_datasets}. 

Either subset fulfills the independence requirement for observations that underlies the statistical model for the forecast error process $V$ in (\ref{likelihood}). 

We consider two surrogate transition densities for process $V$, based on the beta and truncated normal distributions, that lead to approximate log-likelihoods in (\ref{eq:loglikelihoodV-beta}) and (\ref{eq:loglikelihoodV-truncated-gaussian}), respectively.

The inferential strategy aims to recover the unknown diffusion parameters $\bm{\theta} = (\theta_0, \alpha)$ and the threshold parameter $\varepsilon$ for the SDE model (\ref{VtSDE}) at any combination of a subset of the entire data set and proposed surrogate transition density for the process $V$.
This section presents the results of applying the computational method for calibrating the threshold parameter $\varepsilon$ and diffusion parameters $(\theta_0, \alpha)$. 

We start with the ad hoc iterative procedure for calibrating the threshold parameter $\varepsilon$, presented in Appendix~\ref{app_varepsilon}. The iterative procedure repeats until the absolute difference between two consecutive values of $\varepsilon$ becomes smaller than $10^{-3}$. Convergence of the iterative procedure is achieved after at most four iterations for the initial values of $\varepsilon$ (0.002, 0.02, and 0.1) across both data sets and the two proposed surrogate density cases.

The calibrated estimates of the threshold parameter $\varepsilon$ provided by Algorithm~\ref{iterative-process} in Appendix~\ref{app_varepsilon} are as follows:

\begin{table}[H]
\caption{\label{tab:epsilon-beta} Calibrated estimates of $\varepsilon$ for the surrogate beta distribution.}
	\centering
	\noindent\begin{minipage}{\linewidth}
		\centering
		\begin{tabular}{|c|c|c|} 
	\hline
	&    $\hat{\varepsilon}$ & \% of boundary data  \\ 
	\hline
	Data Set 1 & 0.073 & 6.51           \\
	\hline
	Data Set 2 & 0.085 & 6.56  \\
	\hline
\end{tabular}
	\end{minipage}
\end{table}

\begin{table}[H]
\caption{\label{tab:epsilon-truncated-gaussian} Calibrated estimates of $\varepsilon$ for the surrogate truncated normal distribution.}
	\centering
	\noindent\begin{minipage}{\linewidth}
		\centering
		\begin{tabular}{|c|c|c|} 
	\hline
	&    $\hat{\varepsilon}$ & \% of boundary data   \\ 
	\hline
	Data Set 1 & 0.067 & 6.17           \\
	\hline
	Data Set 2 & 0.079 & 6.21       \\
	\hline
\end{tabular}
	\end{minipage}
\end{table}

Next, we apply the $\hat{\varepsilon}$ values to the thresholded normalized forecast function $\left(\frac{p_t}{h_t}\right)_{\!\!\varepsilon}$ in (\ref{corrforecast}) and compute the initial guesses of the diffusion parameters using formulas (\ref{initheta0}) and (\ref{initheta0alpha}) on the entire data sets. Those initial estimates are displayed in Tables~\ref{tab:initial-parameters-beta} and \ref{tab:initial-parameters-truncated-gaussian} below, for the beta and truncated normal cases, respectively.

\begin{table}[H]
\caption{\label{tab:initial-parameters-beta} Initial estimates of the diffusion parameters using the entire data sets with calibrated threshold $\hat{\varepsilon}$ for the surrogate beta distribution.}
	\centering
	\noindent\begin{minipage}{\linewidth}
		\centering
		\begin{tabular}{|l|c|c|c|} 
			\hline
			&    $\theta_0$ & $\alpha$ & $\theta_0\alpha$  \\ 
			\hline
				Data Set 1 & 21.04      & 0.12    &2.52   \\
			\hline
				Data Set 2 & 19.22      & 0.12   &2.31   \\
			\hline
		\end{tabular}
	\end{minipage}
\end{table}

\begin{table}[H]
\caption{\label{tab:initial-parameters-truncated-gaussian} Initial estimates of the diffusion parameters using the entire data sets with calibrated threshold $\hat{\varepsilon}$ for the surrogate truncated normal distribution.}
	\centering
	\noindent\begin{minipage}{\linewidth}
		\centering
		\begin{tabular}{|l|c|c|c|}  
			\hline
			&    $\theta_0$ & $\alpha$ & $\theta_0\alpha$  \\ 
			\hline
				Data Set 1 & 21.10     & 0.12    & 2.53   \\
			\hline
				Data Set 2 & 19.21      & 0.12   & 2.31   \\
			\hline
		\end{tabular}
	\end{minipage}
\end{table}

Given the initial values of the diffusion parameters in Tables~\ref{tab:initial-parameters-beta} and \ref{tab:initial-parameters-truncated-gaussian} and their corresponding $\hat{\varepsilon}$ values in Tables~\ref{tab:epsilon-beta} and \ref{tab:epsilon-truncated-gaussian}, we use the entire data sets to minimize the negative log-likelihoods (\ref{eq:loglikelihoodV-beta}) and (\ref{eq:loglikelihoodV-truncated-gaussian}) over $(\theta_0, \alpha)$. The optimal estimates of the diffusion parameters are displayed next in Tables~\ref{tab:optimal-parameters-beta} and \ref{tab:optimal-parameters-truncated-gaussian}.

\begin{table}[H]
\caption{\label{tab:optimal-parameters-beta} Optimal estimates of the diffusion parameters for the beta distribution on the entire data sets.}
	\centering
	\noindent\begin{minipage}{\linewidth}
		\centering
		\begin{tabular}{|l|c|c|c|} 
			\hline \xrowht{15pt}
&    $\hat{\theta}_0$ & $\hat{\alpha}$ & ${\hat{\theta}}_0 {\hat{\alpha}}$  \\ 
\hline
Data Set 1 &  22.50      & 0.16    &   3.60        \\
\hline
Data Set 2 & 20.63      & 0.16   &   3.30    \\
\hline
		\end{tabular}
	\end{minipage}
\end{table}

\begin{table}[H]
\caption{\label{tab:optimal-parameters-truncated-gaussian} Optimal estimates of the diffusion parameters for the truncated normal distribution on the entire data sets.}
	\centering
	\noindent\begin{minipage}{\linewidth}
		\centering
		\begin{tabular}{|l|c|c|c|} 
			\hline \xrowht{15pt}
&    $\hat{\theta}_0$ & $\hat{\alpha}$ & ${\hat{\theta}}_0 {\hat{\alpha}}$  \\ 
\hline
Data Set 1 &  21.75     & 0.16    &   3.48        \\
\hline
Data Set 2 & 20.95      & 0.16   &   3.35      \\
\hline
		\end{tabular}
	\end{minipage}
\end{table}

The application of the computational procedure for inferring all unknown parameters in the SDE model (\ref{VtSDE}) with beta and truncated normal surrogate distributions is complete. 
To globally check the results, we use all data and optimal estimates derived in each case (combination of both data sets, with the surrogate distribution) to compute the negative log-likelihood level sets as a function of $(\theta_0,\alpha)$ on a fine grid. In each case, the global minimum of the negative log-likelihood function is marked as a purple triangle point. \\

\begin{figure}[H]
	\minipage{0.49\textwidth}%
	\includegraphics[width=\linewidth]{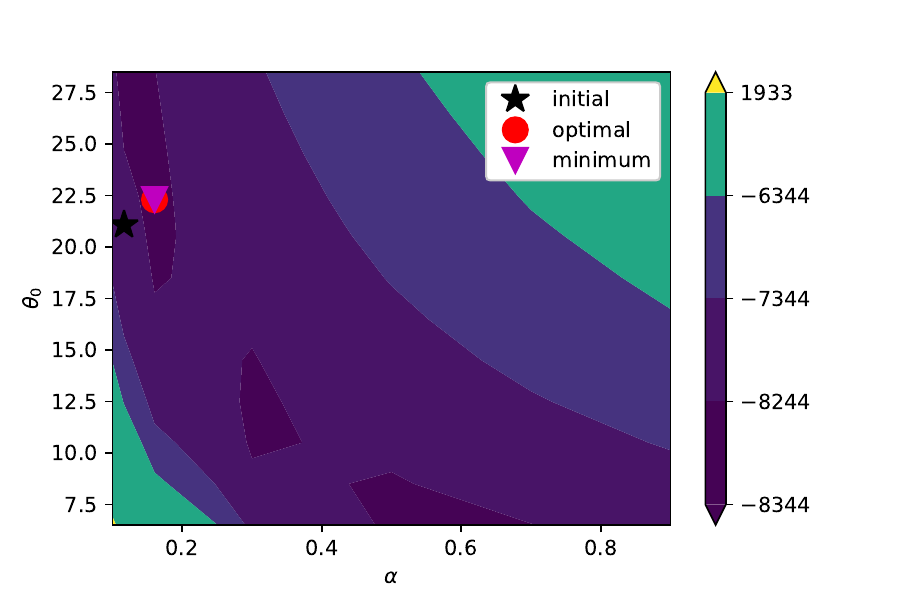}
	\caption*{Data Set 1}
	\label{fig:levelsets-dataset1}
	\endminipage
	\minipage{0.49\textwidth}%
	\includegraphics[width=\linewidth]{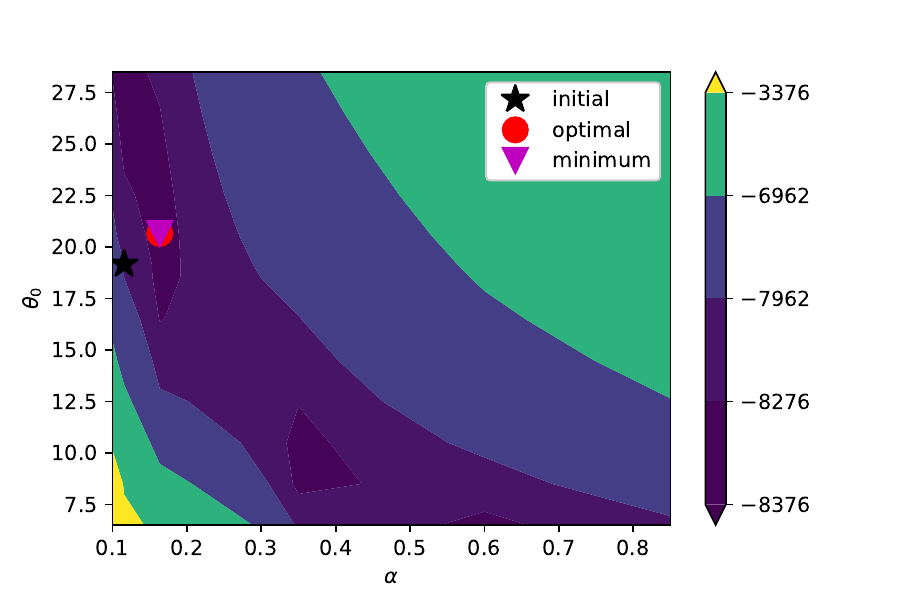}
	\caption*{Data Set 2}
	\label{fig:levelsets-dataset2}
	\endminipage\\
	\caption{Negative log-likelihood level sets as a function of $(\theta_0,\alpha)$ for the beta distribution.} \label{fig:min-beta}
\end{figure}

\begin{figure}[H]	
	\minipage{0.49\textwidth}%
	\includegraphics[width=\linewidth]{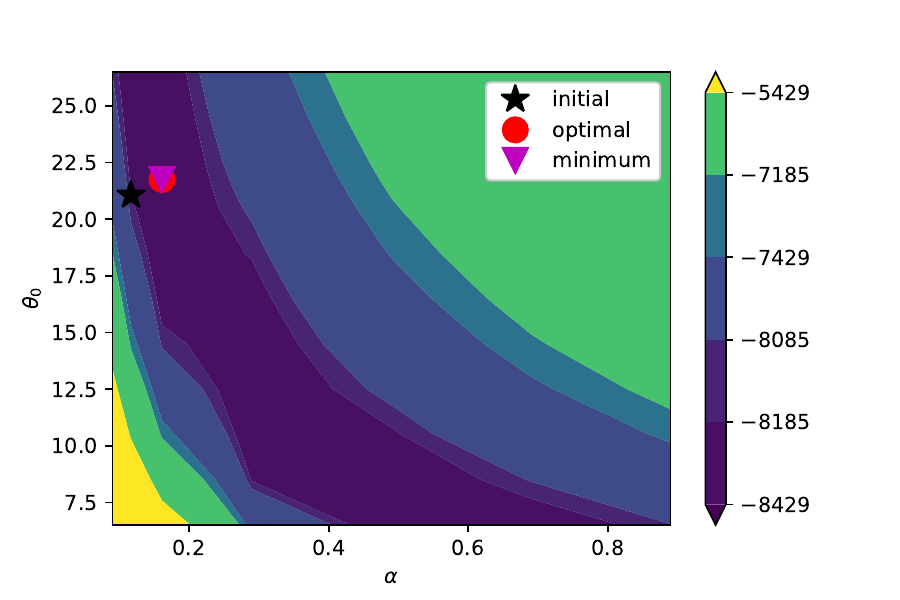}
	\caption*{Data Set 1}
	\label{fig:levelsets_truncated-gauss-dataset1}
	\endminipage
	\minipage{0.49\textwidth}%
	\includegraphics[width=\linewidth]{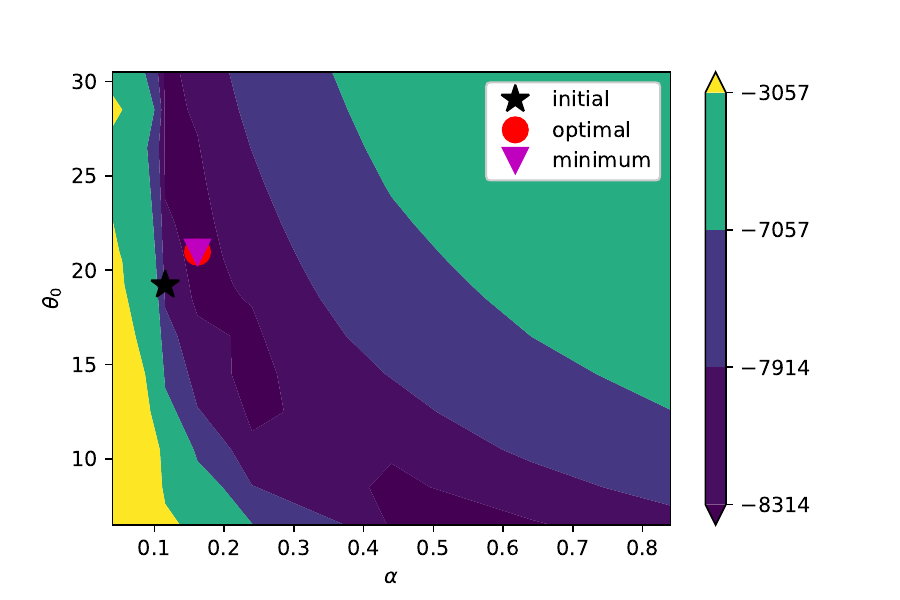}
	\caption*{Data Set 2}
	\label{fig:levelsets_truncated-gauss-dataset2}
	\endminipage
	\caption{Negative log-likelihood level sets as a function of $(\theta_0,\alpha)$ for the truncated normal distribution.}  \label{fig:min-trunc}
\end{figure}

Figures~\ref{fig:min-beta} and  \ref{fig:min-trunc} demonstrate the high matching between the optimal estimates of $(\theta_0, \alpha)$, marked as a red circle in each case (values in Tables~\ref{tab:optimal-parameters-beta} and \ref{tab:optimal-parameters-truncated-gaussian}), and the global minima (purple triangles).


\subsection{Path simulations}

The optimal estimates of the parameters $(\hat{\theta}_0,\hat{\alpha})$ and $\hat{\varepsilon}$ for the SDE model (\ref{VtSDE}) derived in the previous section for the beta distribution and truncated normal distribution are used with the projected Euler scheme to simulate the paths of the normalized solar PV power production $X$ defined in (\ref{ourmodel}). We use the following algorithm to simulate the paths of $X$ with $\Delta=0.0069$. 

\begin{algorithm}[H]
	\caption{Normalized solar photovoltaic power path simulation} 
	\label{solar-path}
	\begin{algorithmic}[1]
	\State \textbf {Input}:  $\hat{\theta}_0$, $\hat{\alpha}$, $\hat{\varepsilon}$, $p=p_{j,i}$, $\dot{p}=\dot{p}_{j,i}$, $h=h_{j,i}$, $j \in \{1,\dots,M\}$, $i \in \{0,\dots, N_j - 1\}$; time spacing $\Delta$ 
		\For {$j=1 \textrm{ to } M$}
				\State \textbf{Initialization}: Set $X_{j,0}=p_{j,0}$;
			\For {$i=0 \textrm{ to }  N_j -1$}
				\State \textbf { Generate } $\Delta W_{i} \sim \mathcal{N}(0,\Delta)$;
	\If {$X_{j,i}<0$}
	\State $X_{j,i}=0$;
	\ElsIf{$X_{j,i}>h$}
	\State $X_{j,i}=h$;
	\EndIf
	\State \textbf { Return }:
	 $X_{j,i+1}=X_{j,i}+(\dot{p}_{j,i}-\theta_i\left(X_{j,i}-p_{j,i}\right))\Delta + \sqrt{2 \alpha \theta_{0} X_{j,i}\left(h_{j,i}-X_{j,i}\right)} \Delta W_{i}$;
	\EndFor
			\EndFor
	\end{algorithmic} 
\end{algorithm}

\begin{figure}[H]
\minipage{0.49\textwidth}
  \includegraphics[width=\linewidth]{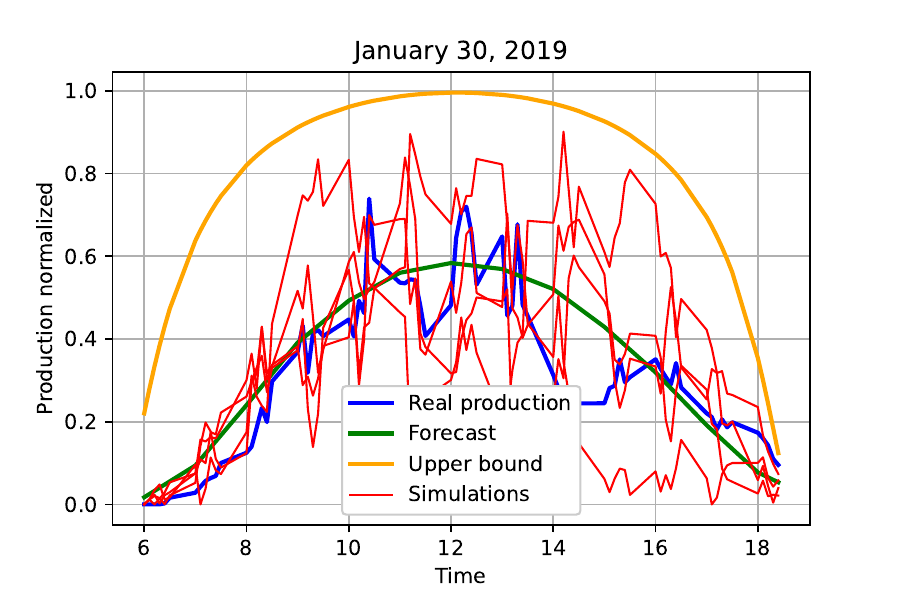}
\endminipage\hfill
\minipage{0.49\textwidth}
  \includegraphics[width=\linewidth]{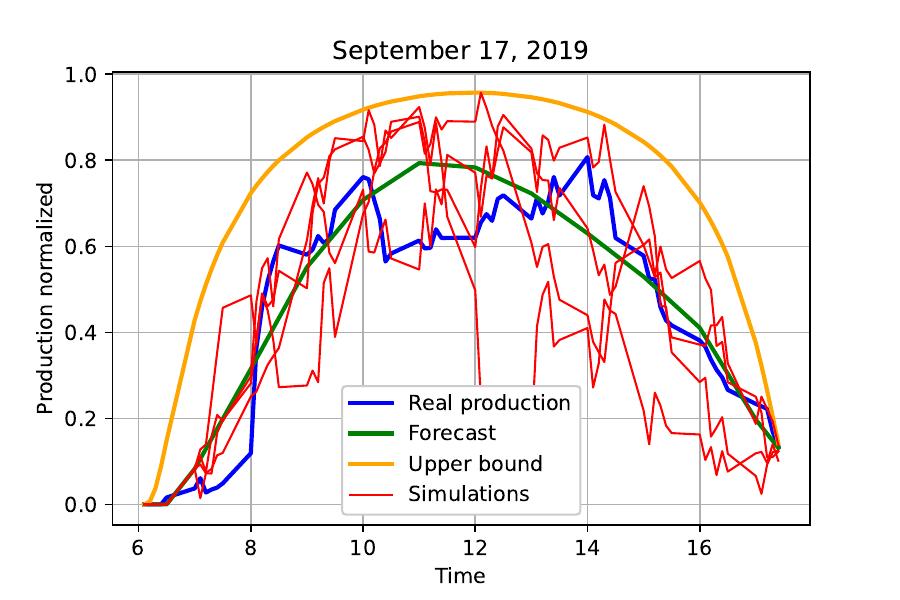}
\endminipage\hfill
\caption{Five simulated normalized solar photovoltaic power production paths with the beta distribution, $(\hat{\theta}_0, \hat{\alpha})=(22.50 , 0.16)$ and $\hat{\varepsilon}=0.073$. }
\label{fig:path-beta-dataset2}
\end{figure}

\begin{figure}[H]
	\minipage{0.49\textwidth}
	\includegraphics[width=\linewidth]{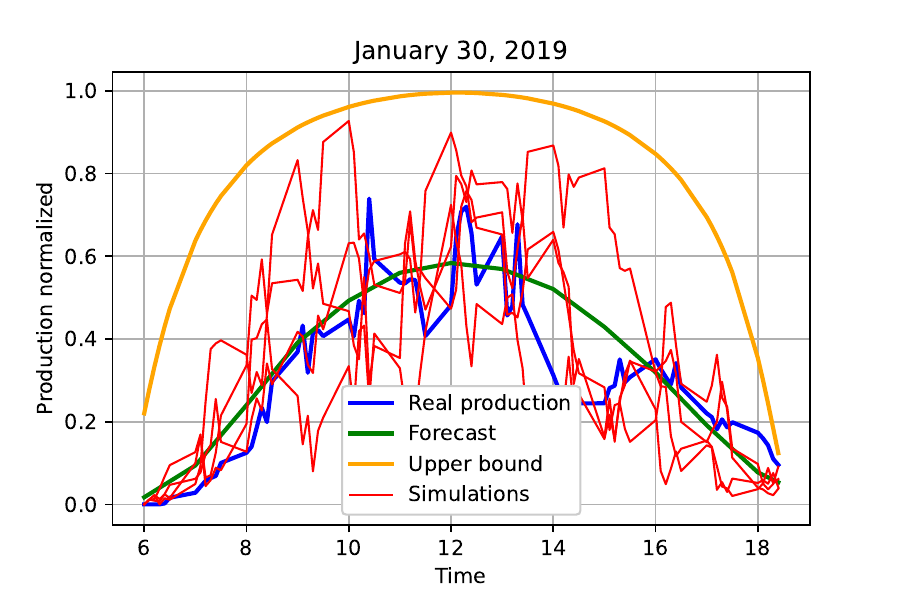}
	\endminipage\hfill
	\minipage{0.49\textwidth}
	\includegraphics[width=\linewidth]{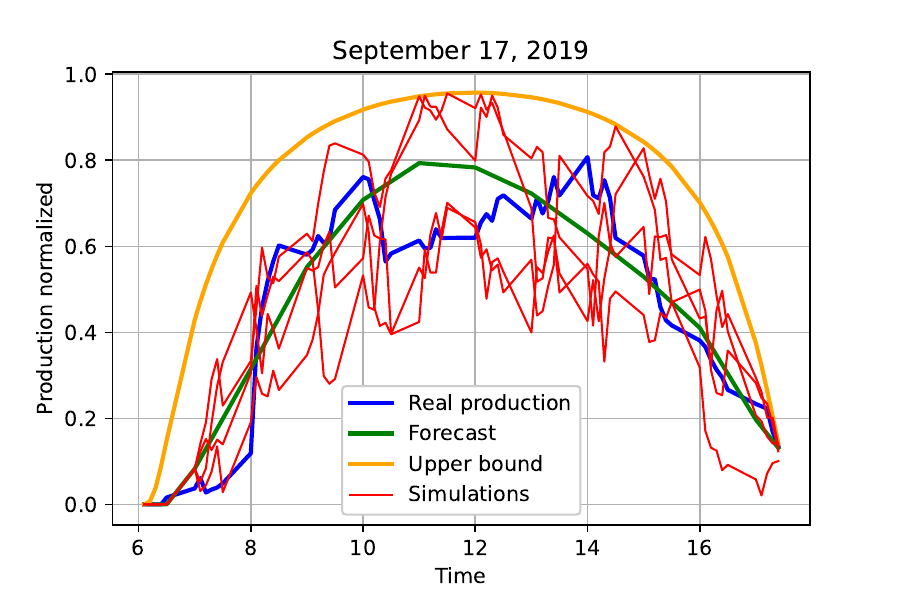}
	\endminipage\hfill
	\caption{Five simulated normalized solar photovoltaic power production paths with the truncated normal distribution $(\hat{\theta}_0, \hat{\alpha})=(21.75, 0.16)$ and $\hat{\varepsilon}=0.067$. }
	\label{fig:path-truncgaussian-dataset2}
\end{figure}

Figures~\ref{fig:path-beta-dataset2} and \ref{fig:path-truncgaussian-dataset2} exhibit five simulations of the normalized solar photovoltaic power production (red lines), normalized actual production (blue line), forecast (green line), and computed upper bound (orange line) for two days in 2019.

The capability to generate path simulations for the normalized solar PV power production using a well-calibrated data-driven constrained SDE model constitutes a valuable tool for electric system operators.

\subsection{Pathwise confidence bands}
\begin{figure}[H]
\minipage{0.49\textwidth}
  \includegraphics[width=\linewidth]{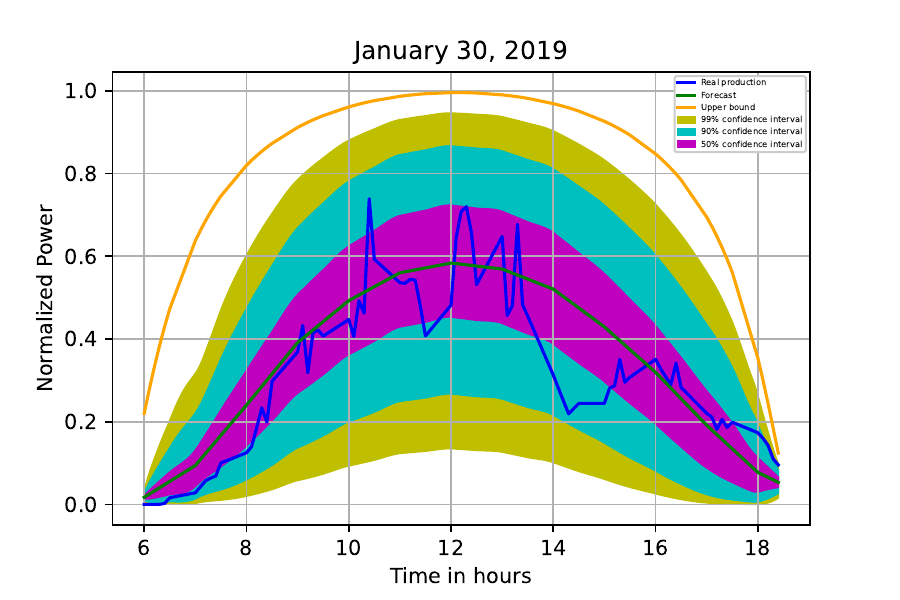}
\endminipage\hfill
\minipage{0.49\textwidth}
  \includegraphics[width=\linewidth]{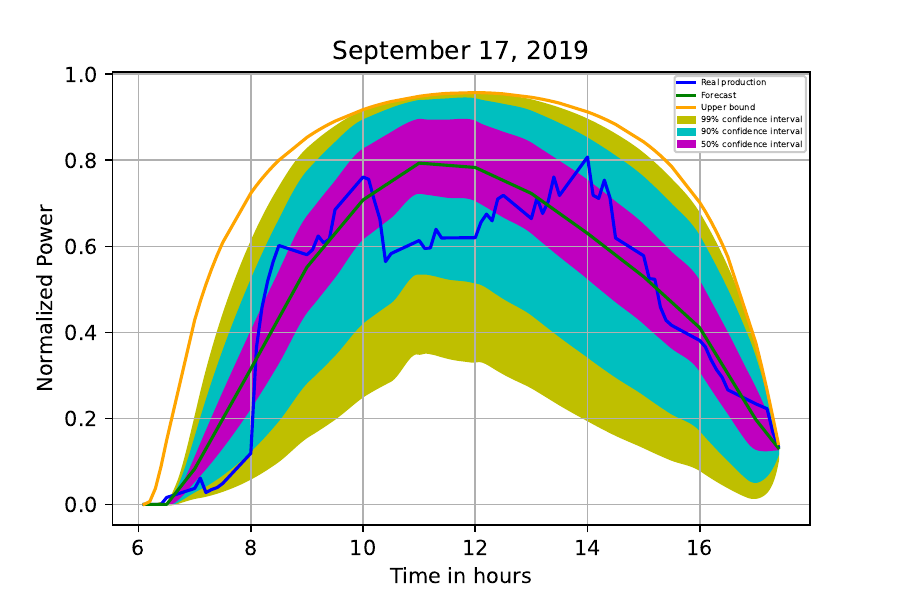}
\endminipage\hfill
\caption{Pathwise confidence bands for solar photovoltaic power production using the approximate maximum likelihood estimates for the beta proxy $(\hat{\theta}_0, \hat{\alpha})=(22.50, 0.16)$ and $\hat{\varepsilon}=0.073$. Blue line: real production, green line: forecast, orange line: computed solar upper bound.}
\label{fig:band-beta-dataset2}
\end{figure}

\begin{figure}[H]
	\minipage{0.49\textwidth}
	\includegraphics[width=\linewidth]{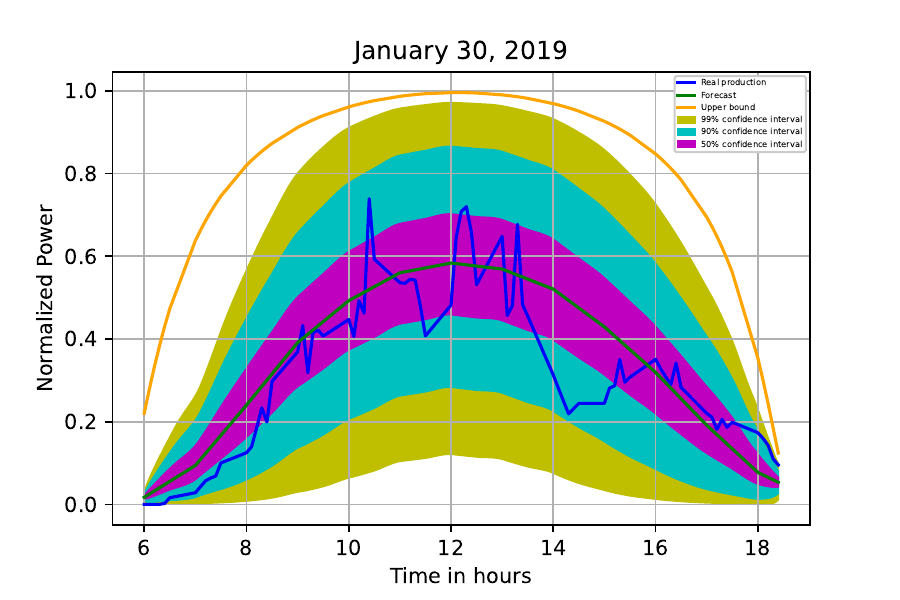}
	\endminipage\hfill
	\minipage{0.49\textwidth}
	\includegraphics[width=\linewidth]{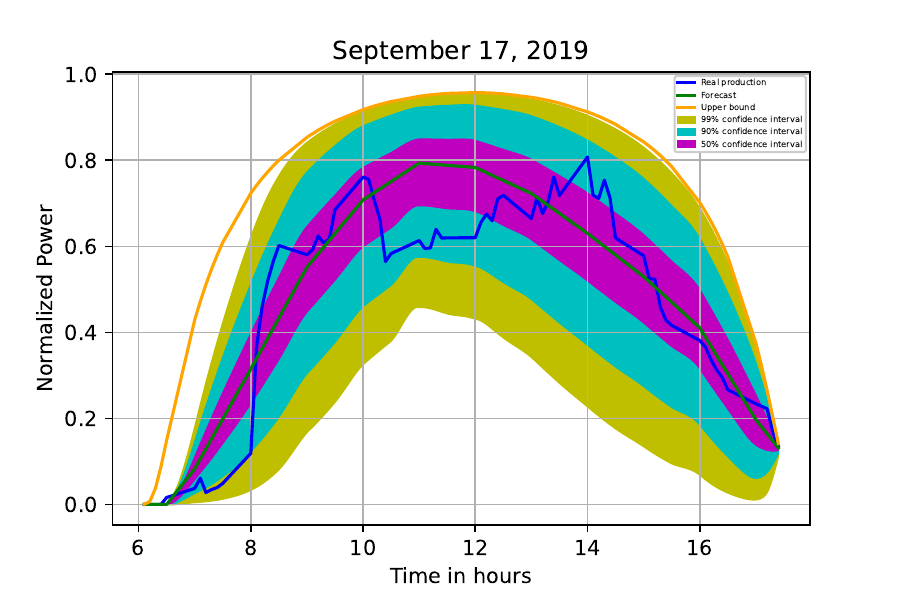}
	\endminipage\hfill
	\caption{Pathwise confidence bands for solar photovoltaic power production using the approximate maximum likelihood estimates for the truncated normal proxy $(\hat{\theta}_0, \hat{\alpha})=(21.75, 0.16)$ and $\hat{\varepsilon}=0.067$. Blue line: real production, green line: forecast, orange line: computed solar upper bound.}
	\label{fig:band-truncated-gaussian-dataset2}
\end{figure}

Pathwise confidence bands with 99\%, 90\%, and 50\% prescribed coverage are obtained through the following indirect inference technique. For each day, we compute the moments using the theoretical moment equations. Then, we employ the link equations between the moments and parameters that characterize the selected statistical approximation of the transition density, enabling the computation of the theoretical quantiles of the surrogate distribution. The results of these computations calculated using the daily segments of Data Set 2, using the optimal estimates of the parameters found by calibrating the SDE model (\ref{VtSDE}) with Data Set 1, are presented in Figures~\ref{fig:band-beta-dataset2} and \ref{fig:band-truncated-gaussian-dataset2}.

\subsection{Log-likelihood comparison of statistical approximations of the SDE model  (\ref{VtSDE})}

The evaluation of the transition density of the process $V$ (the solution to the SDE model (\ref{VtSDE})) at any day $j$ and time instant $i$ enables comparing the calibration procedure performance at the optimal parametric points for the two selected statistical approximations of the model (\ref{VtSDE}). The comparison is based on the beta density (\ref{eq:loglikelihoodV-beta}) and truncated normal density (\ref{eq:loglikelihoodV-truncated-gaussian}) on the log-likelihood scale. 

For this purpose, we applied the innovative nonparametric kernel smoothing procedure with the control variate method for the pointwise evaluation of the transition density of the process $V$, described in Algorithm~\ref{KDE}.

\begin{figure}[H]
	\minipage{0.49\textwidth}
	\includegraphics[width=\linewidth]{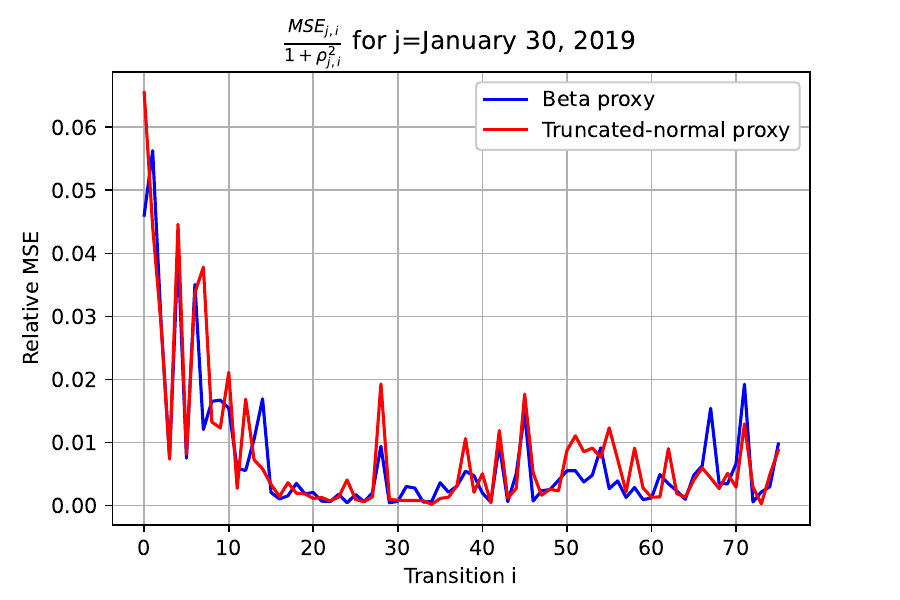}
	\endminipage\hfill
	\minipage{0.49\textwidth}
	\includegraphics[width=\linewidth]{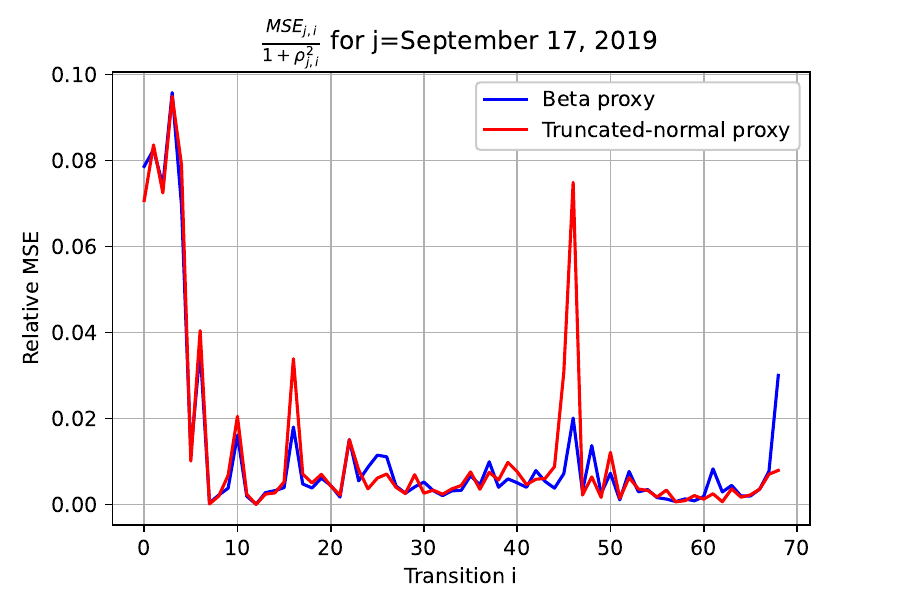}
	\endminipage
	\caption{Relative mean squared error with the beta and truncated normal distributions, $(\hat{\theta}_0, \hat{\alpha})=(22.50, 0.16)$ and $\hat{\varepsilon}=0.073$, and $(\hat{\theta}_0, \hat{\alpha})=(21.75, 0.16)$ and $\hat{\varepsilon}=0.067$.}
	\label{fig:relativeMSE}
\end{figure}

Figure~\ref{fig:relativeMSE} displays the values of the relative MSE for two daily segments of Data Set 2, using the optimal estimates of the parameters found by calibrating the SDE model (\ref{VtSDE}) with the Data Set 1 with the surrogate beta and truncated normal distributions.

In the numerical comparison with the proposed method, the tolerance for the relative MSE reaches below $10\%$, needing fewer simulations than the crude kernel density estimation.

Using the estimator (\ref{kde}), we computed the log-likelihood value that corresponds to any combination of subsets of data and optimal estimates of the model parameters (see Tables~\ref{tab:epsilon-beta}, \ref{tab:epsilon-truncated-gaussian}, \ref{tab:optimal-parameters-beta}, and \ref{tab:optimal-parameters-truncated-gaussian}). The results are summarized in the following table.
\begin{table}[H]
\caption{Log-likelihood values of the stochastic differential equation model (\ref{VtSDE}) computed using the nonparametric estimator (\ref{kde}) with optimal parametric points $({\hat{\varepsilon}}, {\hat{\theta}}_0, {\hat{\alpha}})$ obtained for the truncated normal density as a proxy of the transition density of the forecast process $V$ (Tables~\ref{tab:epsilon-truncated-gaussian} and \ref{tab:optimal-parameters-truncated-gaussian}) and for the beta density proxy (Tables~\ref{tab:epsilon-beta} and \ref{tab:optimal-parameters-beta}) for Data Sets 1 and 2.}
	\label{Likelihood-kde}
	\centering
{\renewcommand{\arraystretch}{1.5}%
	\begin{tabular}{|l|l|l|}
		\hline
		& Data Set 1 & Data Set 2 \\ \hline
		Optimal parametric points $({\hat{\varepsilon}}, {\hat{\theta}}_0, {\hat{\alpha}})$ from Tables~\ref{tab:epsilon-truncated-gaussian} and \ref{tab:optimal-parameters-truncated-gaussian}	& 8593   & 8377     \\ \hline
		Optimal parametric points $({\hat{\varepsilon}}, {\hat{\theta}}_0, {\hat{\alpha}})$ from Tables~\ref{tab:epsilon-beta} and \ref{tab:optimal-parameters-beta}      & 8554   &  8384   \\ \hline
	\end{tabular}}
\end{table}

The property of independence among the days within both data sets allows using the central limit theorem in the case of the sum of independent and not identically distributed random variables to construct asymptotic confidence intervals for the unknown log-likelihood value.

\begin{table}[H]
\caption{Asymptotic 95\% confidence intervals for the log-likelihood value based on the estimated log-likelihood values in Table~\ref{Likelihood-kde} computed by applying Algorithm~\ref{KDE}.}
	\label{Likelihood-kde-confidence-interval} 
	\centering
{\renewcommand{\arraystretch}{1.5}%
	\begin{tabular}{|l|cc|cc|}
		\hline
		\multirow{3}{*}{} & \multicolumn{2}{c|}{Data Set 1}     & \multicolumn{2}{c|}{Data Set 2}                 \\ \cline{2-5} 
& \multicolumn{1}{c|}{Lower} & Upper & \multicolumn{1}{c|}{Lower} & Upper \\[-2ex]
		& \multicolumn{1}{c|}{bound} & bound & \multicolumn{1}{c|}{bound} & bound \\ \hline
		Optimal parametric points  $({\hat{\varepsilon}}, {\hat{\theta}}_0, {\hat{\alpha}})$ from Tables~\ref{tab:epsilon-truncated-gaussian} and \ref{tab:optimal-parameters-truncated-gaussian}	  & \multicolumn{1}{c|}{8392}  & 8793        & \multicolumn{1}{c|}{8182}       & 8572   \\ \hline 
		Optimal parametric points  $({\hat{\varepsilon}}, {\hat{\theta}}_0, {\hat{\alpha}})$ from Tables~\ref{tab:epsilon-beta} and \ref{tab:optimal-parameters-beta}                   & \multicolumn{1}{c|}{8356}        & 8751        & \multicolumn{1}{c|}{8188}       & 8580        \\ \hline
	\end{tabular}}
\end{table}

From a practical viewpoint, the above procedure provides a reliable criterion for choosing an adequate density proxy candidate that better fits the actual data with a low time cost. Moreover, the efficiency of Algorithm~\ref{KDE} for the computation of the estimator (\ref{kde}) allows for determining the log-likelihood value of the SDE model (\ref{VtSDE}) at any point of interest in the parametric space.



\section{Conclusions} \label{Section_5}

This paper developed a generalization of the proposed framework (\hcite{2021ckst}) based on It\^{o}'s SDEs to assess the pathwise uncertainty of the forecast error along a short-term horizon. 

From the probabilistic modeling perspective, the novel feature in the SDE model is a time-dependent upper boundary in the drift and diffusion functions, limiting the normalized nonnegative observable phenomenon of interest and its forecast. The time-dependent upper function represents valuable information that can be a) estimated from physics-based data, b) provided by field-knowledge experts, or c) recognized as external input from other sources. We adjusted the rigorous mathematical theory developed in our previous work (\hcite{2021ckst}) to the present framework with the constrained data. Thus, the forecast error process is modeled through an SDE whose linear drift term has a time-varying mean-reversion parameter depending upon the forecast and upper bound functions, their derivatives, and unknown parameters. The nonlinear state-dependent Jacobi-type diffusion term features the normalized thresholded forecast function and unknown parameters. The model is designed to prevent the paths of the forecast error process from reaching the time-dependent boundaries due to a condition on the time-varying mean-reversion parameter. Truncating the normalized forecast function through a threshold parameter ensures that the time-varying mean-reversion parameter remains bounded.

We introduced innovative statistical inference methods to address the proposed SDE model, whose transition density is unavailable in a closed analytical form. We calibrated the SDE model to approximate the transition density of the forecast error process selected by the user. The calibration focuses first on the threshold parameter, applying a fast ad hoc iterative procedure, then optimizing the approximate log-likelihood to estimate the other SDE model parameters.
The likelihood value of the original SDE model is computed through a novel nonparametric point estimation algorithm of the transition density. We coupled another SDE model to the original one with the same drift function. The diffusion function of the auxiliary model is expressed through a time-dependent parameter that we estimated using the moment-matching technique. We used simulations from the coupled system of SDEs to compute a nonparametric point estimator of the transition density based on a tailored kernel smoothing technique with the control variate method. The number of simulations was chosen adaptively to ensure the relative MSE is below the desired threshold. This procedure allows the relative performance of the calibration technique at the optimal parametric points obtained for the statistical versions of the SDE model specified by the user to be checked efficiently on the log-likelihood scale.

The new methodology is illustrated through a case study to characterize the pathwise uncertainty of the forecast error of the 2019 daily solar PV power production in Uruguay, with the actual available power production every 10 min and the one-day ahead forecast data every hour. In this study, the time-dependent upper bound is given by the daily maximum solar PV power production, which we compute using the ``very simple clear sky model'' (\hcite{sandiaGHI}, pp.~14-15) to determine the irradiance function. 

Future work can develop methods to directly optimize the likelihood function of the SDE model without considering statistical approximations for its transition density. Moreover, the forecast function could be updated as new data become available. It could be explored the performance of SDE models with a generalized diffusion coefficient, for example, in the form proposed by \hciteauthor{bggk}, considering $ \sqrt{2 \alpha \theta_0} X_t^{\beta} (h_t - X_t)^{\gamma}$, $\frac{1}{2} \leq \beta, \gamma \leq 1$.
 


\begin{appendices}

\section{Ad hoc iterative procedure for the calibration of the threshold parameter}\label{app_varepsilon}

The ad hoc iterative procedure we propose for the automatic calibration of the threshold parameter $\varepsilon$ is based on splitting the entire sample of transitions of $V$ into two disjoint subsets, identified as follows:
\begin{itemize}
\item the ``inner'' data set that contains all transitions such that $\varepsilon_{\text{init}} < \frac{p_t}{h_t} < 1 -  \varepsilon_{\text{init}}$; 
\item the ``boundary'' data set that contains all transitions such that $\frac{p_t}{h_t} \geq 1 -  \varepsilon_{\text{init}}$ or $ \frac{p_t}{h_t} \leq \varepsilon_{\text{init}}$, 
\end{itemize}
where $\varepsilon_{\text{init}}$ is an initial value for $\varepsilon$ defined by the user. \\
Given $\varepsilon_{\text{init}}$, we can compute the initial estimates for the baseline mean-reversion parameter $\theta_0$ and the diffusion variability coefficient $\theta_0 \alpha$ using only the inner data set. \\
 
Remark: As an initial guess for the baseline mean-reversion parameter $\theta_0$, after solving a regression problem for the conditional mean of $V$, as in the work by \hciteauthor{2021ckst}, except when using only the inner data set, we obtain 

\begin{equation}
\theta_0^* \approx\frac{ \mathop{\sum_{j=1}^M \sum_{i=1}^{N_j}}\limits_{(i,j): \varepsilon_{\text{init}} < \frac{p_{j,i}}{h_{j,i}} < 1 -  \varepsilon_{\text{init}}}  v_{j,i-1}(v_{j,i-1}-v_{j,i})}{\Delta \cdot \mathop{\sum_{j=1}^M \sum_{i=1}^{N_j}}\limits_{(i,j): \varepsilon_{\text{init}} < \frac{p_{j,i}}{h_{j,i}} < 1 -  \varepsilon_{\text{init}}}  (v_{j,i-1})^2}. 
\label{initheta0}
\end{equation}

The quadratic variation of $V$ 
\begin{equation}
\int_0^t b(V_s; \bm{\theta}, p_s, h_s)^2 ds = 
\int_0^t 2 \alpha \theta_0 \left(V_s + \frac{p_s}{h_s} \right) \left(1 - V_s - \frac{p_s}{h_s}\right)  ds
\end{equation}
is approximated by the discrete sum $\sum_{0< t_{j, i-1} \leq t}\left(V_{t_{j, i}} - V_{t_{j, i-1}}\right)^2$. As an initial guess for the diffusion variability coefficient $\theta_0 \alpha$, we obtain the following:
\begin{equation}
(\theta_0 \alpha)^*=\frac{ \mathop{\sum_{j=1}^M \sum_{i=1}^{N_j}}\limits_{(i,j): \varepsilon_{\text{init}} < \frac{p_{j,i}}{h_{j,i}} < 1 -  \varepsilon_{\text{init}}} (v_{j,i} - v_{j,i-1})^2}{2\Delta\cdot \mathop{\sum_{j=1}^M \sum_{i=1}^{N_j}}\limits_{(i,j): \varepsilon_{\text{init}} < \frac{p_{j,i}}{h_{j,i}} < 1 -  \varepsilon_{\text{init}}} (v_{j,i}+ \frac{p_{j,i}}{h_{j,i}})(1- v_{j,i} - \frac{p_{j,i}}{h_{j,i}})},  \label{initheta0alpha}
\end{equation}
where $\Delta$ is the length of the time interval between two consecutive measurements. \\
The first stage of the algorithm optimizes the approximate log-likelihood over $(\theta_0, \alpha)$, again restricted to the inner data set. The second stage of the algorithm, with the updated estimates of $(\theta_0, \alpha)$, optimizes the approximate log-likelihood over $\varepsilon$ using only the boundary data set. The two stages iterate until the convergence of the $\varepsilon$ estimate reaches the desired accuracy. The goal of this iterative two-stage ad hoc procedure, whose pseudocode is summarized below, is to provide a calibrated estimate for the threshold parameter $\varepsilon$.

\begin{algorithm}[H]
	\caption{Ad hoc iterative calibration procedure for the threshold parameter $\varepsilon$} 
	\label{iterative-process}
	\begin{algorithmic}[1]
		\State \textbf {Input}:  $\varepsilon_{\text{init}}$, $\frac{p}{h}= \frac{p_{j,i}}{h_{j,i}}$, $j \in {1,\dots,M}$, $i \in {0,\dots,N_j}$
		\State  \textbf{Divide}
		\begin{itemize}
			\item ``inner'': $\varepsilon_{\text{init}} < \frac{p}{h} < 1 -  \varepsilon_{\text{init}}$; 
			\item ``boundary'': $\frac{p}{h} \geq 1 -  \varepsilon_{\text{init}}$ or $ \frac{p}{h} \leq \varepsilon_{\text{init}}$. 
		\end{itemize}
		\State \textbf {Compute} $\theta_{0\text{init}}$ and $\alpha_{\text{init}}$ using inner with (\ref{initheta0}) and (\ref{initheta0alpha})
		\State $\left(\theta_{0},\alpha\right)\gets$\textbf {Optimize} (\ref{eq:loglikelihoodV-beta}) or (\ref{eq:loglikelihoodV-truncated-gaussian}) over $\left(\theta_{0},\alpha\right)$ using ``inner''
		\State $\varepsilon\gets$\textbf{Optimize} (\ref{eq:loglikelihoodV-beta}) or (\ref{eq:loglikelihoodV-truncated-gaussian}) over $\varepsilon$ using ``boundary''
		\State  \textbf{while} {$|\varepsilon-\varepsilon_{\text{init}}|>10^{-3}\:$ \textbf{do}}\\
		$\varepsilon_{\text{init}}\gets \varepsilon$
		\State  \textbf{Divide}
		\begin{itemize}
			\item ``inner'': $\varepsilon_{\text{init}} < \frac{p}{h} < 1 -  \varepsilon_{\text{init}}$; 
			\item ``boundary'': $\frac{p}{h} \geq 1 -  \varepsilon_{\text{init}}$ or $ \frac{p}{h} \leq \varepsilon_{\text{init}}$.
		\end{itemize}
		\State \textbf {Compute} $\theta_{0\text{init}}$ and $\alpha_{\text{init}}$ using ``inner'' with (\ref{initheta0}) and (\ref{initheta0alpha})
		\State $\left(\theta_{0},\alpha\right)\gets$\textbf {Optimize} (\ref{eq:loglikelihoodV-beta}) or (\ref{eq:loglikelihoodV-truncated-gaussian}) over $\left(\theta_{0},\alpha\right)$ using ``inner''
		\State $\varepsilon\gets$\textbf{Optimize} (\ref{eq:loglikelihoodV-beta}) or (\ref{eq:loglikelihoodV-truncated-gaussian}) over $\varepsilon$ using ``boundary''
		\State  \textbf{end while}
		\State \textbf {Output}: $\hat{\varepsilon}$, calibrated estimate of $\varepsilon$
	\end{algorithmic} 
\end{algorithm}

\section{Bias variance analysis of a kernel density estimator with control variate}\label{app_varbound}
\begin{Thm}\label{thm:msekde}
Let $V$ denote a real random variable with density $\rho_V$. For $v\in\mathbb R$,  let $\hat \rho_m(v;h)$ denote the following kernel density estimator of $\rho_V(v)$ with control variate given by 
\begin{equation}
	{\widehat{\rho}}_m(v; h) :=\frac{1}{m}\sum_{n=1}^{m}\left(\kappa_{h}(V^{(n)}-v)-\kappa_{h}(Z^{(n)}-v)\right) + \mathbb{E} \left[\kappa_{h}\left(Z- v\right)\right],
\end{equation}
where $x\mapsto \kappa_{h}(x)=\frac{1}{\sqrt{2\pi}h}\exp\left(-\frac{x^2}{2h^2}\right)$ denotes the Gaussian kernel with bandwidth $h>0$, and $Z$ a Gaussian random variable with mean $\mu_Z$  and variance $\sigma^2_Z$.
\begin{enumerate}
\item If the density of $V$ has a bounded second derivative $\rho^{\prime\prime}_V$, then  for all $v\in\mathbb R$
$$|\mathbb E\hat \rho_m(v;h) - \rho_V(v)|\le \frac{h^2\|\rho^{\prime\prime}_V\|_{\infty}}{2}.$$
\item If  the random variable $V$ is with bounded density such that $\mathbb E[V^4]<\infty$, then for all $v\in\mathbb R$
\begin{equation*}
\text{Var}({\widehat{\rho}}_m(v; h)) \le  \frac 1{\pi h^{\frac 72}m}{\mathbb E}^{\frac 12}\left[ \Bigl| (V-v)^2 - (Z-v)^2 \Bigr|^2\right] 
\sqrt{ \left(\sqrt{\pi}\|\rho_V\|_{\infty}+ \frac{1}{\sqrt{2} \sigma_Z}\right)}.
\end{equation*}
\end{enumerate}
\end{Thm}
\begin{proof} 
For the first assertion,  we use a change of variable and a Taylor expansion to write
\begin{align*}
\mathbb E[\hat \rho_m(v;h)] - \rho_V(v)&=\int_{\mathbb R}\Big[ \rho_V(zh+v)-\rho_V(v)\Big]\frac{e^{-\frac{z^2}{2}}}{\sqrt{2\pi}}dz\\
&=\frac{h^2}2\int_{\mathbb R}\Big[\int_0^1\rho^{\prime\prime}_V(v+\lambda zh) d\lambda\Big] z^2\frac{e^{-\frac{z^2}{2}}}{\sqrt{2\pi}}dz\\
&\le \frac{h^2\|\rho^{\prime\prime}_V\|_{\infty}}{2}.
\end{align*}
For the second assertion, we use that $|e^a - e^b|^2\le 2|e^a -e^b|$ for  $a,b<0$, to get
\begin{align*}
\text{Var}({\widehat{\rho}}_m(v; h))&\le \frac 1{2\pi h^2m} \mathbb E\left|\exp\left(-\frac{(V-v)^2}{2h^2}\right) -\exp\left(-\frac{(Z-v)^2}{2h^2}\right) \right|^2\\
&\le \frac 1{\pi h^2m} \mathbb E\left|\exp\left(-\frac{(V-v)^2}{2h^2}\right) -\exp\left(-\frac{(Z-v)^2}{2h^2}\right) \right|\\
&\le  \frac 1{2\pi h^4m}\mathbb E\left[ \Bigl| (V-v)^2 - (Z-v)^2 \Bigr| \left( \exp\left(-\frac{(V-v)^2}{2h^2}\right) + \exp\left(-\frac{(Z-v)^2}{2h^2}\right)\right)\right],
\end{align*}
where we used that $|e^a-e^b|\le |a-b|e^{a\vee b}\le |a-b|(e^a+ e^b)$, for all $a,b\in \mathbb R$.
Then, by the Cauchy-Schwarz inequality and the elementary inequality $|a-b|^2\le 2(a^2+b^2)$, we get 
\begin{align*}
\text{Var}({\widehat{\rho}}_m(v; h))&\le  \frac 1{\pi h^4m}{\mathbb E}^{\frac 12}\left[ \Bigl| (V-v)^2 - (Z-v)^2 \Bigr|^2  \right] \sqrt{{\mathbb E}\left[ \exp\left(-\frac{(V-v)^2}{h^2}\right) +\exp\left(-\frac{(Z-v)^2}{h^2}\right)  \right] }.
\end{align*}
We obtain the result by noticing that 
$$
\left |{\mathbb E}\left[ \exp\left(-\frac{(V-v)^2}{h^2}\right) \right]\right |=h\int_{\mathbb R}\exp(-x^2)\rho_{V}(hx+v)dx \le h\sqrt{\pi}\|\rho_V\|_{\infty}
$$
and
$$
\left |{\mathbb E}\left[ \exp\left(-\frac{(Z-v)^2}{h^2}\right) \right]\right |=h\int_{\mathbb R}\exp(-x^2) \frac{\exp(-\frac{(hx+v-\mu_Z)^2}{2\sigma^2_Z})}{\sqrt{2\pi}\sigma_Z}dx\le \frac{h}{\sqrt{2}\sigma_Z}.$$
\end{proof}

\section{Computation of the irradiance function $I_D(\cdot)$ and daily maximum solar photovoltaic power production $h(\cdot)$}
\label{app_irr}

We recall that $I_{D}(\cdot) \,[\text{MW/m}^2]$, the direct normal irradiance as a function of time is given by the expression (\ref{eq:irrf}):
\begin{equation*}
 I_{D}(\cdot)= \frac{ 1353 \times 0.7^{AM_r(\cdot)^{0.678}}}{1000}\,,
\end{equation*}
where $1353\,\text{W/m}^2$ is an estimate of the solar constant (\hcite{dbb20}, pp.~5--6; \hcite{SAYIGH1979}), and $AM_r(\cdot)$ is the approximate empirical expression of the relative optical air mass that considers the refraction effect of the atmosphere on solar radiation propagation (\hcite{IQBAL198385}, p.~100), originally derived by \hciteauthor{Kasten}, as a function of the elevation angle  $\alpha \,\text{[deg]}$:
\begin{equation*}
AM_r(\cdot) =  [\sin(\alpha) + 0.15 (\alpha + 3.885)^{-1,253}]^{-1} \,.
\end{equation*}
In the previous expression, for simplicity, we omitted that the elevation angle (also known as the solar altitude, related to the zenith angle $\theta_z = 90^{\circ} - \alpha$) depends explicitly on time. Such dependence is emphasized below. The elevation angle, representing the angle between the horizon (horizontal surface) and the line to the Sun measured from the horizon throughout the day, is given by \hcite{Karathanasis2019}, p.~47:
\begin{equation*}
        \alpha=\arcsin( \sin(\phi_s)\, \sin(\delta) + \cos(\phi_s)\, \cos(\delta)\, \cos(\omega))\,,
\end{equation*}
where $\phi_s \,\text{[deg]}$ indicates the solar site latitude, $\delta \,\text{[deg]}$ denotes the declination angle, and $\omega \,\text{[deg]}$ represents the hour angle (also called the solar time angle because it is just an angular representation of the solar time). Several formulas are available for the Sun's declination $\delta$ (\hcite{Karathanasis2019}, p.~49), which depends only on time and not a geographical location. We used 
\begin{equation*}
   \delta= 23.45^{\circ} \sin \left(360^{\circ} \frac{(284 +d)}{365} \right) = - 23.45^{\circ} \sin \left(360^{\circ} \frac{(81 -d)}{365} \right) = - 23.45^{\circ} \cos \left(360^{\circ} \frac{(10.25 + d)}{365} \right)\,,
\end{equation*}
where $d$ is the day number ($d=1$ for January~1). \\
The hour angle $\omega$ is the angular displacement of the Sun east or west of the local meridian due to rotation of the Earth on its axis at $15^{\circ}$ per hour (\hcite{sandiaGHI}, p.~12; \hcite{Karathanasis2019}, p.~13), which is computed as
\begin{equation*}
    \omega = 15^{\circ} \left[  \underbrace{\left( LT + \overbrace{\frac{4^{\prime}}{60} \frac{(LSTM - L_s)}{1^{\circ}} + \frac{EoT}{60}}^{\text{timescale correction}} \right)}_{\text{local solar time [hour]}} - 12 \text{ hours}  \right]\,,
\end{equation*}
where $LT \equiv t\,[\text{hour}]$ denotes the local standard clock time, $L STM\,\text{[deg]} := 15^{\circ}\, \Delta_{GMT}$ represents the local standard time meridian, $\Delta_{GMT}$ indicates the difference in the local time ($LT$) from the Greenwich mean time ($GMT$) in hours, and $L_s \,\text{[deg]}$ denotes the longitude of the solar site. Moreover, $EoT \,\text{[min]}$, the ``equation of time'' (\hcite{sandiaGHI}, p.~11; \hcite{Karathanasis2019}, p.~53), is an empirical equation that corrects for the eccentricity of the Earth's orbit around the Sun and the Earth's axial tilt with respect to the ecliptic plane:
\begin{equation*}
EoT=9.87 \sin(2B) - 7.53 \cos(B) - 1.5 \sin(B)\,, \quad \text{ with} \quad B=360^{\circ}\frac{(d - 81)}{365}
\end{equation*}
where $B \,\text{[deg]}$ denotes the day angle, and $d$ represents the day number. \\

Once given the irradiance function $I_D(\cdot)$, we compute the constant $k\,[\text{m}^2]$ by dividing the maximum installed solar PV power capacity in Uruguay during 2019 provided by UTE, using 228.8~MW for the computed irradiance and taking the maximum of these ratios. We found the constant $k=257057~\text{m}^2$. Finally, the time-dependent upper bound in this case study (i.e., the daily maximum solar PV power production $h(t)$) is computed as $h(t) = k\, I_D(t)$ throughout 2019.

\section{Data Sets 1 and 2: Graphical data analysis}
\label{app_datasets}

\begin{figure}[H]
	\minipage{0.49\textwidth}
	\includegraphics[width=\linewidth]{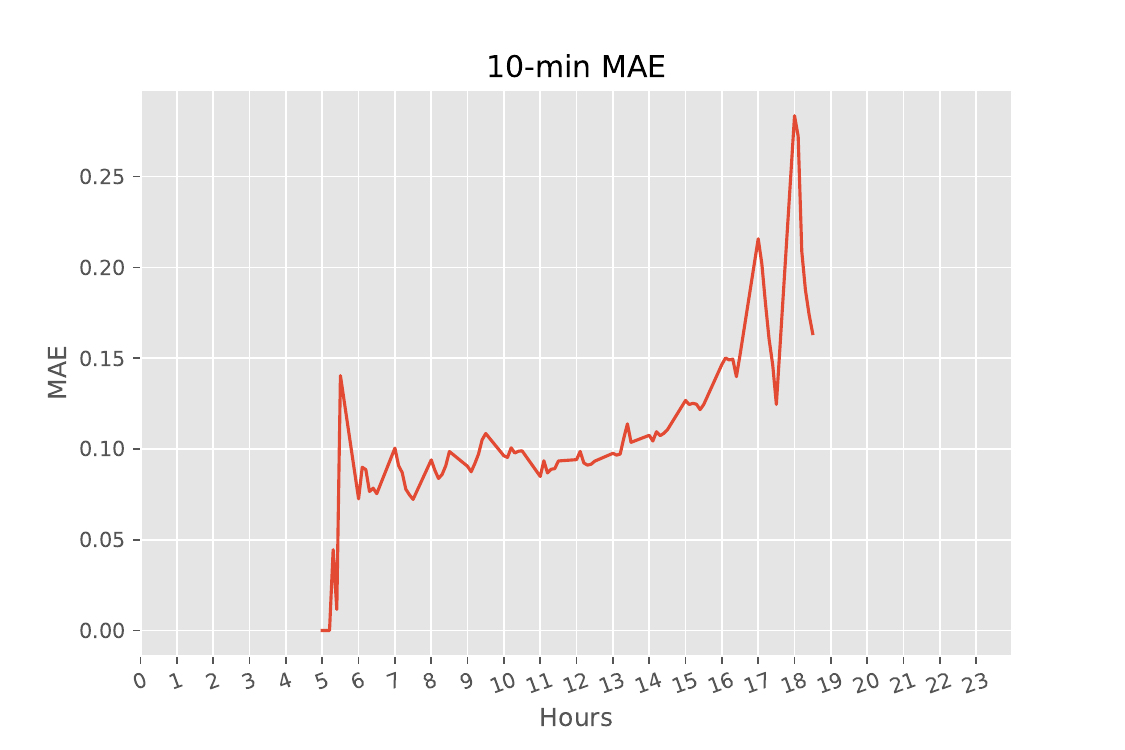}
	\caption*{Ten-min mean absolute error (MAE).}\label{fig:hourly_mae_datase1}
	\endminipage\hfill
	\minipage{0.49\textwidth}
	\includegraphics[width=\linewidth]{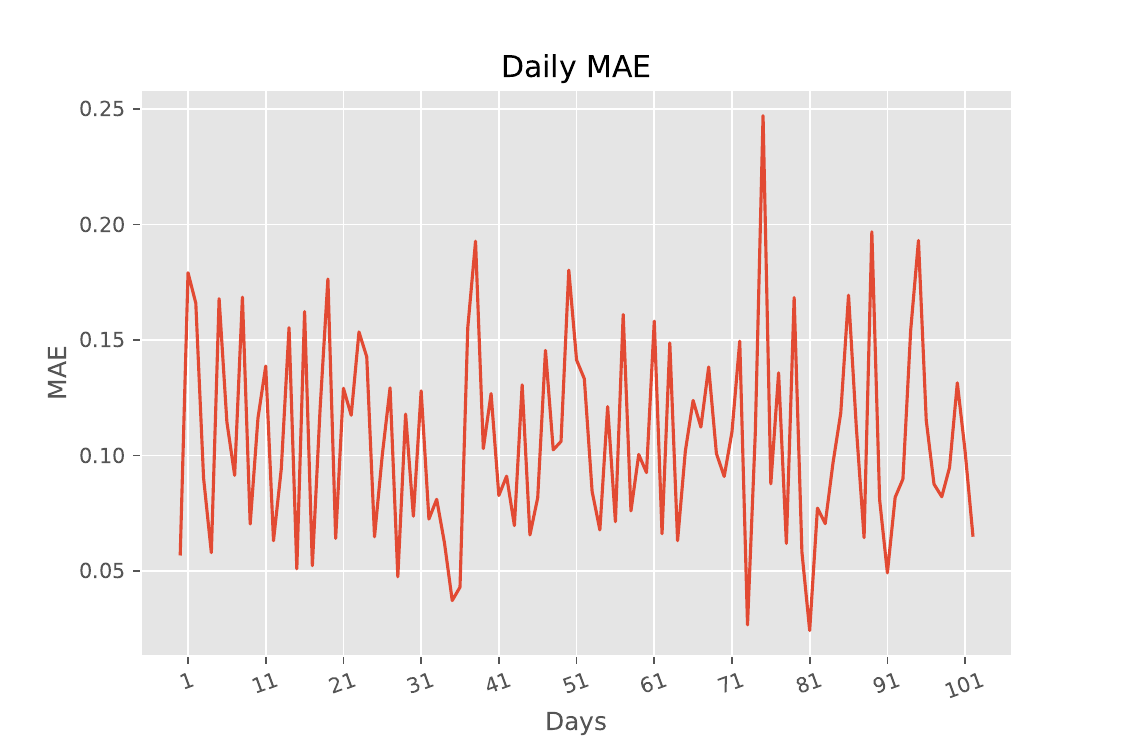}
	\caption*{Daily mean absolute error (MAE).}\label{fig:daily_mae_datest1}
	\endminipage
	\caption{Data Set 1.}
\end{figure}

\begin{figure}[H]
	\minipage{0.49\textwidth}
	\includegraphics[width=\linewidth]{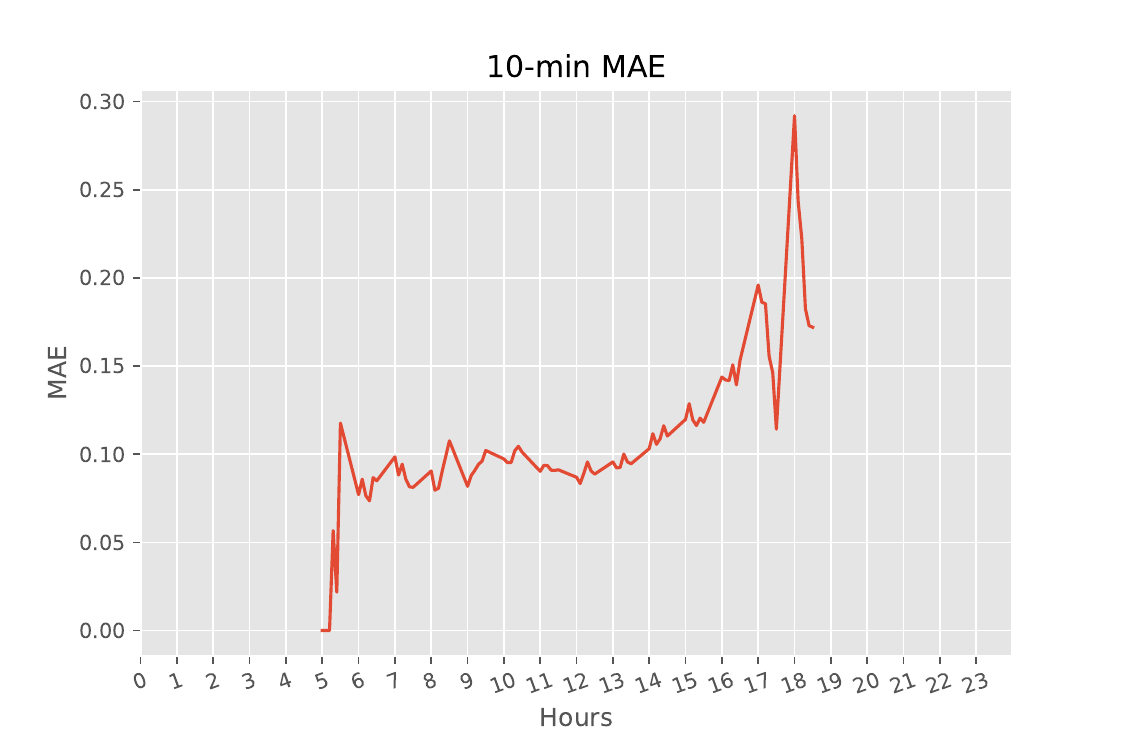}
	\caption*{Ten-min mean absolute error (MAE).}\label{fig:hourly_mae_datase2}
	\endminipage\hfill
	\minipage{0.49\textwidth}
	\includegraphics[width=\linewidth]{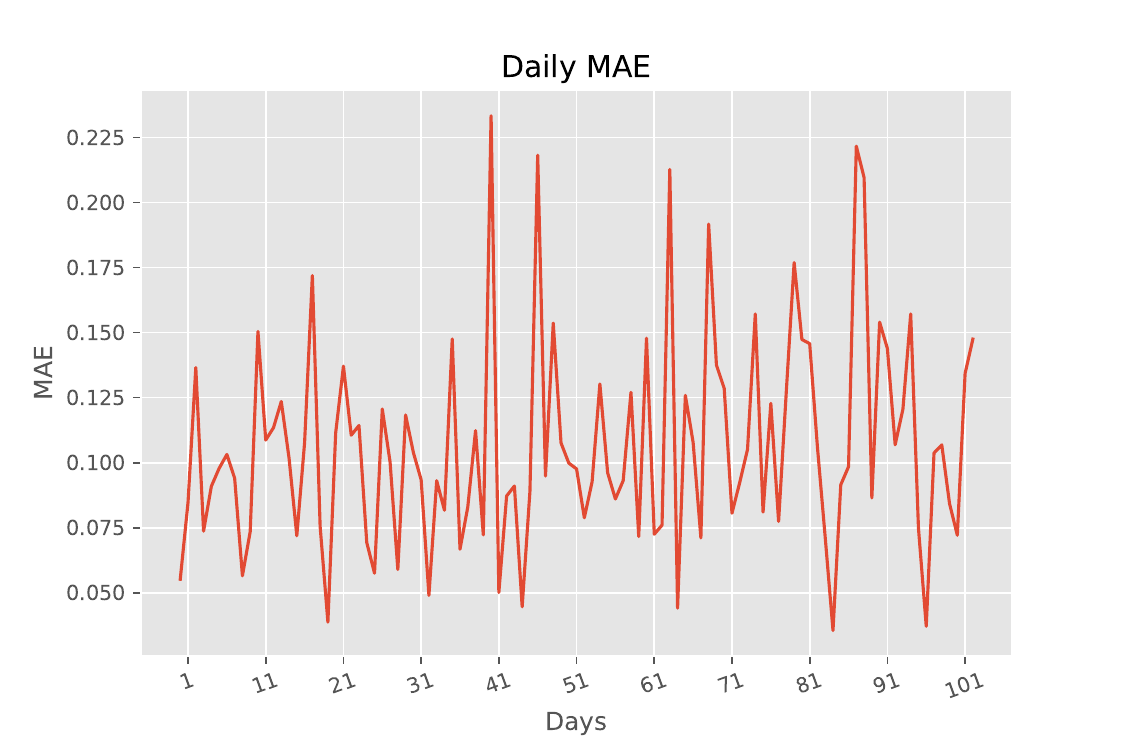}
	\caption*{Daily mean absolute error (MAE).}\label{fig:daily_mae_datest2}
	\endminipage
	\caption{Data Set 2.}
\end{figure}

\begin{figure}[H]
    \centering
   \minipage{0.58\textwidth}%
  \includegraphics[width=\linewidth, height=6cm]{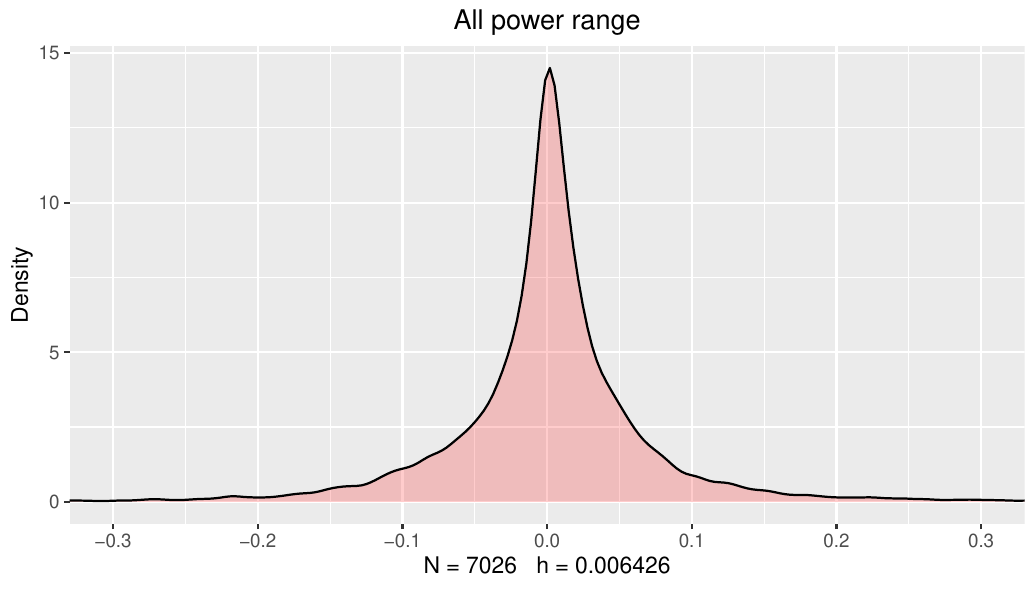}
  \caption{Kernel density estimation based on the solar photovoltaic power forecast error transitions for Data Set 1 with standard normal kernel and Scott's bandwidth selection method.}
  \label{fig:error_forecast_all} 
\endminipage
\end{figure}

\begin{figure}[H]
    \centering
   \minipage{0.58\textwidth}%
  \includegraphics[width=\linewidth,  height=6cm]{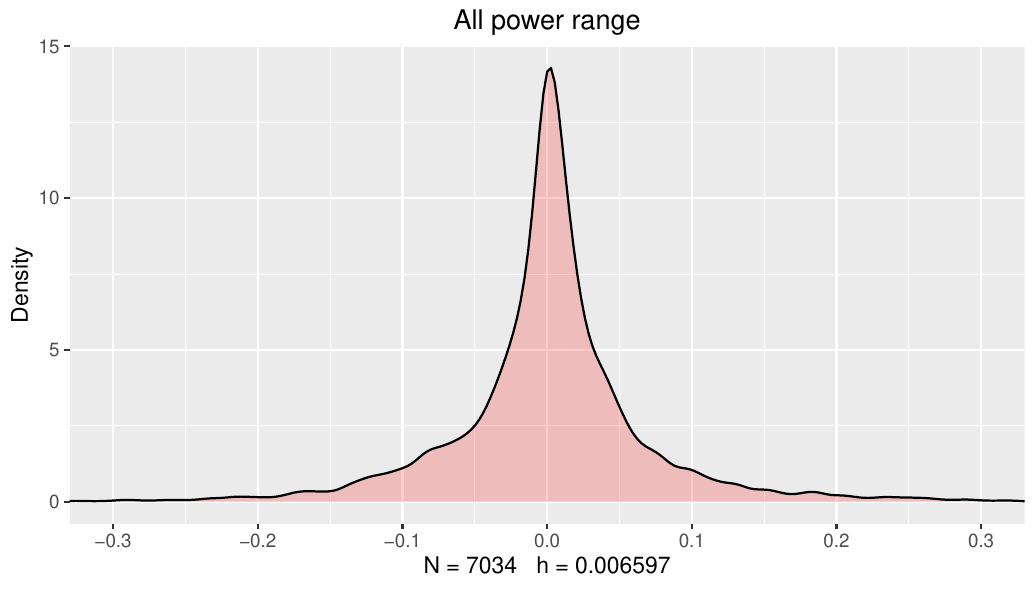}
  \caption{Kernel density estimation based on the solar photovoltaic power forecast error transitions for Data Set 2 with standard normal kernel and Scott's bandwidth selection method.}
  \label{fig:error_forecast_all}
\endminipage
\end{figure}

\begin{figure}[H]
	\centering
	\minipage{0.7\textwidth}%
	\includegraphics[width=\linewidth]{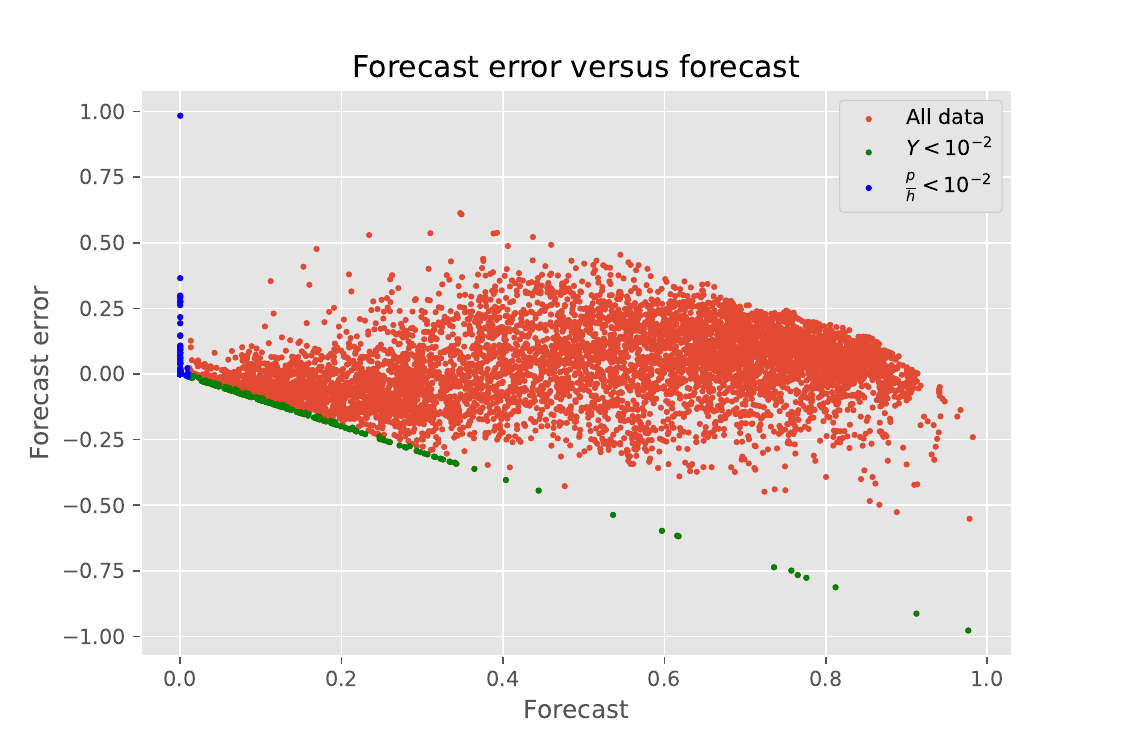}
	\label{fig:error_forecast_g-dataset1}
	\endminipage
	\caption{Data Set 1: forecast error $v_t$ versus forecast $p_t$.}
\end{figure}
\begin{figure}[H]
	\centering
	\minipage{0.7\textwidth}%
	\includegraphics[width=\linewidth]{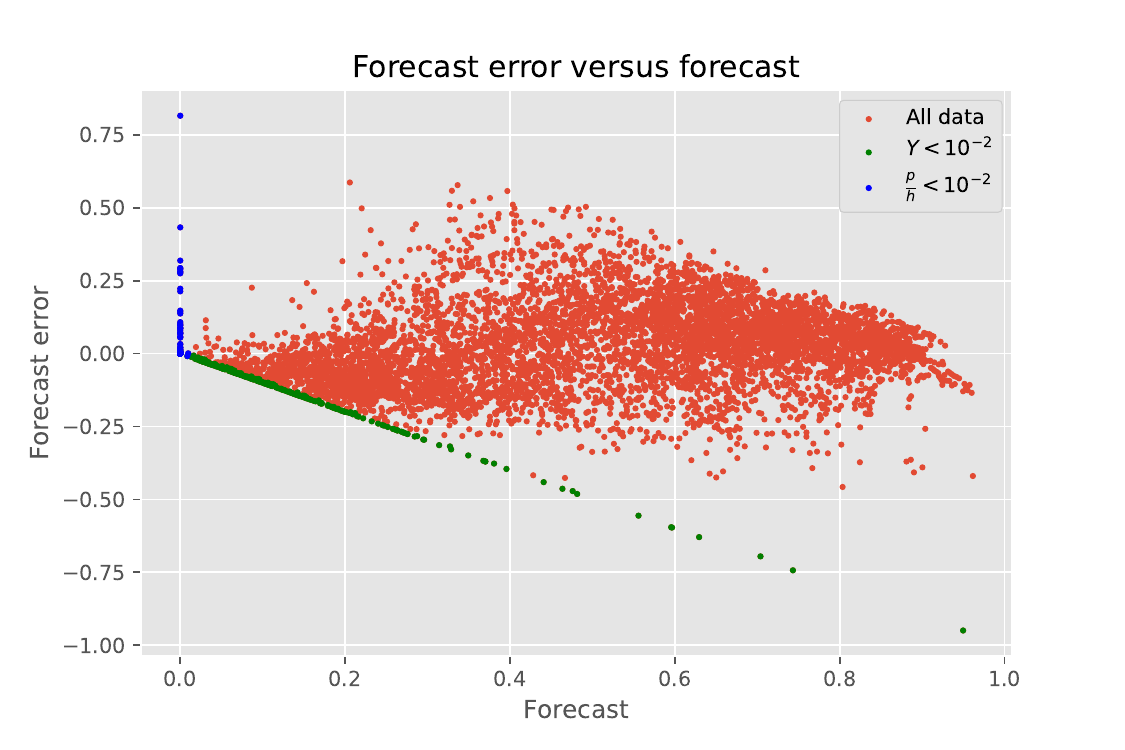}
	\label{fig:error_forecast_g-dataset2}
	\endminipage
	\caption{Data Set 2: forecast error $v_t$ versus forecast $p_t$.}
\end{figure}

\end{appendices}

\subsection*{Acknowledgments} 

This research was partially supported by the KAUST Office of Sponsored Research (OSR) under Award number URF/1/2584--01--01 in the KAUST Competitive Research Grants Program Round 8, the Alexander von Humboldt Foundation, the chair Risques Financiers, Fondation du Risque, and the Laboratory of Excellence MME-DII Grant no. ANR11-LBX--0023--01 (\url{http://labex-mme-dii.u-cergy.fr/}). We thank UTE (\url{https://portal.ute.com.uy/}) for providing the data used in this research.


\nocite{*}
 
\printbibliography[title={References}]

\end{document}